\documentclass[intlimits,twoside,a4paper]{article}
\usepackage{amsmath,amssymb}
\usepackage{graphicx}
\usepackage{wrapfig}
\usepackage[T2A]{fontenc}
\usepackage[cp1251]{inputenc}
\usepackage[eqsecnum]{cmpj}
\issue{2011}{14}{3}{33603}

\doinumber{10.5488/CMP.14.33603}

\title[Simulations of chain-like polyelectrolyte]{Correlation between flexibility of chain-like polyelectrolyte and
thermodynamic properties of its solution}

\author{T.~Sajevic, J.~Re\v{s}\v{c}i\v{c}, V.~Vlachy}
\address{Faculty of Chemistry and Chemical Technology, University of
Ljubljana, \\A\v{s}ker\v{c}eva 5, SI--1000 Ljubljana}

\authorcopyright{T.~Sajevic, J.~Re\v{s}\v{c}i\v{c}, V.~Vlachy, 2011}

\date{Received May 19, 2011, in final form June 15, 2011}

\begin{document}
\maketitle
\begin{abstract}
Structural and thermodynamic properties of the model solution
containing charged oli\-go\-mers and the equivalent number of
counterions were studied by means of the canonical Monte Carlo
simulation technique. The oligomers are represented as (flexible)
freely jointed chains or as a linear (rigid) array of charged hard
spheres. In accordance with the primitive model of electrolyte
solutions, the counterions are modeled as charged hard spheres and
the solvent as dielectric continuum. Significant differences in the pair
distribution functions, obtained for the rigid (rod-like) and flexible
model are found but the differences in thermodynamic properties,
such as, enthalpy of dilution and excess chemical potential, are less
significant. The results are discussed in light of the experimental data
an aqueous polyelectrolyte solutions. The simulations suggest that
deviations from the fully extended (rod-like) conformation yield
slightly stronger binding of counterions. On the other hand, the
flexibility of polyions, even when coupled with the ion-size effects,
cannot be blamed for qualitative differences between the
theoretical results and experimental data for enthalpy of dilution.


\pacs 61.20.Ja, 82.35.Rs, 87.10.Rt, 87.15.A-

\end{abstract}

\section{Introduction}

In more than 50 years of intensive polyelectrolyte research a large
number of experimental and theoretical papers have contributed
toward better understanding of these solutions (for review, see~\cite{alexandrowicz,katchalsky,dolar,dautz,schmitz,forster,springer,dobrynin1,wandrey,blaul,Holm2004}).
Nature and synthetic chemistry have provided polyelectrolytes of
different shapes: they can be rod-like as, for example, DNA, flexible
(chain-like) as are many of the synthetic polyelectrolytes.  Their
conformation in the aqueous solution depends on the chemical
structure of the polyion, presence of low-molecular electrolyte, and
polyelectrolyte concentration. Polyelectrolytes containing carboxylic
groups, for example, poly(acrylic acid) or poly(methacrylic acid) can
dramatically change their conformation in water; the polyion chain
opens from a coiled to an extended configuration as the chain is
neutralized. In addition, these solutions are characterized by strong
intermolecular association~\cite{jerman}.

Fully sulfonated poly(styrenesulfonic) acid and its salts represent a
different class of polyelectrolytes. Due to the nature of sulfonic
group, these (chain-like) polyions assume locally extended
conformation in aqueous solution. Their properties are most often
analyzed within the framework of the rod-like cell model~\cite{alexandrowicz,katchalsky,dolar} or Manning's condensation
approach~\cite{manning0,manning1}. Both models are relatively
simple, treating polyions as infinitely long, uniformly charged
cylinders, or infinite lines. In addition to these simplifications, the
system is treated as continuous dielectric, while interactions other than Coulomb
 are ignored. Despite severe shortcomings, these
purely electrostatic theories yield important insights into the
counterion-polyion interaction.

The basic parameter of classical polyelectrolyte theories is the linear
charge density parameter $\lambda$, defined as
\begin{equation}
\lambda =  \frac{e_{0}^{2}}{4\pi\epsilon_{0}\epsilon_{\mathrm{r}} k_{\mathrm{B}}T b} = \frac{L_{\mathrm{B}}}{b }\,.
\label{lam}\end{equation}

In equation~\eqref{lam}, $e_{0}$ denotes an elementary charge, $b$ is the
length of the monomer unit carrying $e_{0}$, $\epsilon_{\mathrm{r}}$ is relative
permittivity of the system and as usual, $k_{\mathrm{B}}T$ is a product of
Boltzmann's constant and absolute temperature. There is a number of
examples showing\cite{dolar,vesnaver0,arh,lipar,nagaya,blaul} that in
order to bring the Poisson-Boltzmann cell model calculation into
agreement with experimental data  for a given polyelectrolyte,
$\lambda_{\mathrm{eff}}$ should be introduced ($\lambda_{\mathrm{eff}} > \lambda$).
For polystyrenesulfonic acid and its alkaline salts good agreement
between the rod-like cell model calculations and the experimental
data for the osmotic coefficient was obtained for $\lambda_{\mathrm{eff}}
\approx$ 3.4, instead of the structural value~\cite{dolar}, which is most
often taken to be 2.80. This holds true for other
physico-chemical properties~\cite{arh,nagaya,Luksic1,Luksic2,Luksic3}. Although this fact is known
quite some time, the explanation of its origins is unclear.

\begin{figure}[ht]
\begin{center}
\includegraphics[width=8cm]{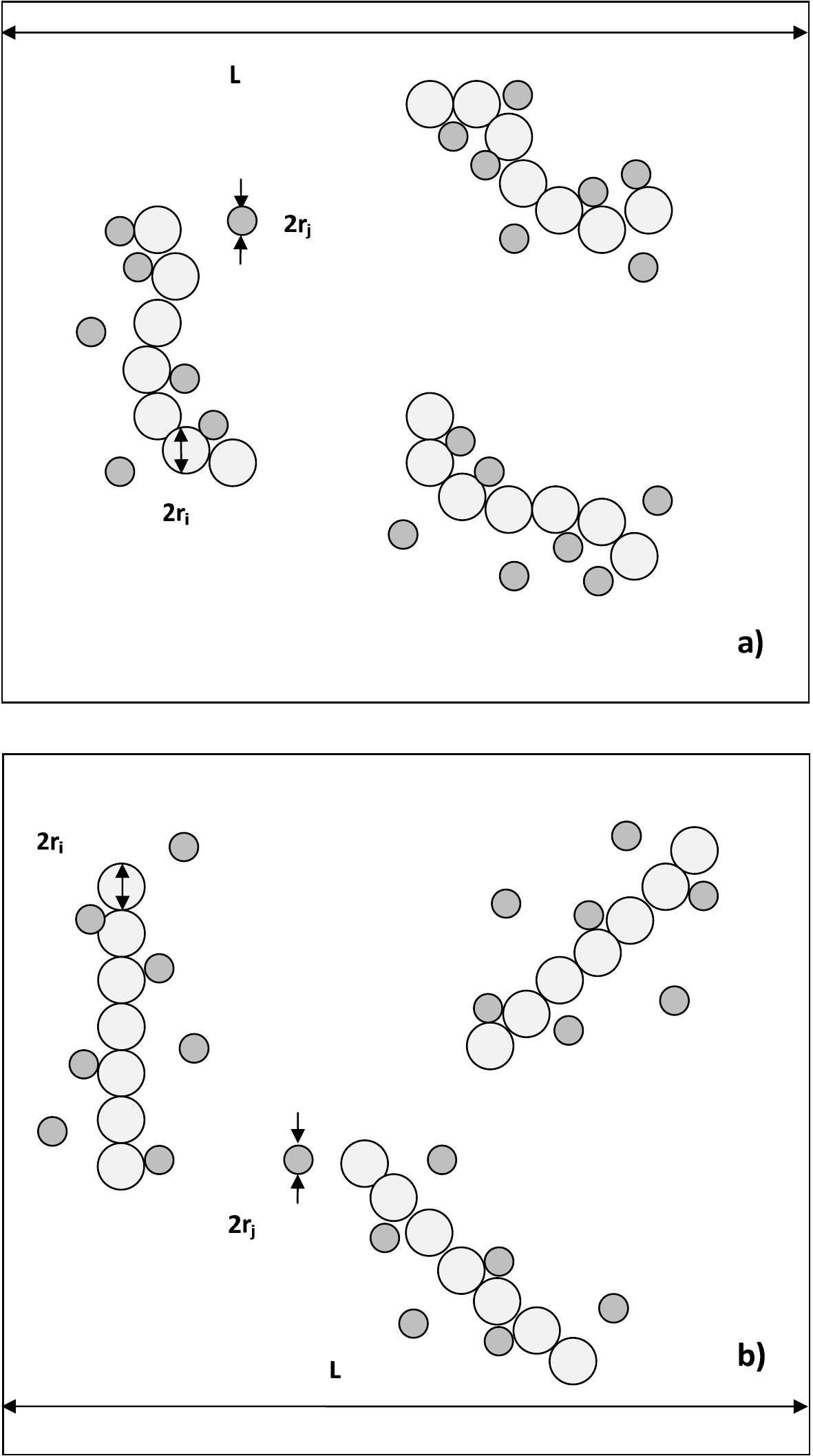}
\caption{Models used in simulation: flexible chains (a) and rigid chains (b).
 $2r_i$ and $2r_j$ are diameters of a monomer unit and a counterion, respectively.}
\label{model}
\end{center}
\end{figure}

One of the reasons often blamed for the disagreement described
above is the fact that the polyions are treated as rigid cylinders, while
realistic polyions are, depending on their molecular structure, more or
less flexible. It is quite clear that a more advanced modeling of
polyelectrolyte solutions should treat polyions as relatively
flexible macromolecules that can adopt various conformations in
solution. Seminal studies of short flexible polyelectrolytes with explicit
counterions are due to Stevens and Kremer~\cite{kremer2,kremer3},
who published molecular dynamics simulations of polyelectrolyte
chains in solution. They used a freely jointed bead-chain polymer
model to describe flexible polyions embedded in dielectric continuum;
the sizes of monomers and counterions were given by repulsive
Lennard-Jones potential. More recently other studies based on the
similar model have been published~\cite{osm,bizjak,antypov}. This
polyelectrolyte model explicitly handles long-range Coulomb
interactions between the charges in solution and those fixed on the
polymer backbone. In contrast to the older studies, which treat finite
polyions as fully extended rigid cylinders~\cite{VD,BD}, new works
explore different conformations and correlate the charge ``binding''
and polymer conformations. Another important study is the work by
Antypov and Holm~\cite{antypov}. These authors performed
simulations of short rod-like polyions for the purpose of examining the
most critical approximations of the rod-like cell model and the
Poisson-Boltzmann theory.

Here we also have to mention an interesting recent contribution by
Carrillo and Dobrynin~\cite{Carrillo}, concerned with another class
of polyelectrolytes in solution. These authors performed a study of
hydrophobic (partially sulfonated) polyelectrolytes, simulating the
transition between the extended and collapsed
state~\cite{Aseyev,Essafi}. This is an extreme case, which does not
apply to highly charged and locally extended polyelectrolytes~\cite{dolar,vesnaver0,arh,lipar,nagaya,Luksic1,Luksic2,Luksic3}, and
is therefore not considered here. The solution properties of such
partially sulfonated polyelectrolytes substantially differ from those we
wish to model here.

In the present contribution we wish to address the effect of flexibility
(departures from fully extended conformation) of a polyion chain on
 thermodynamic properties of solution. In order to get some
quantitative information about the effect, we utilized the Monte Carlo
method to explore two polyelectrolyte models, schematically shown in
figure~1. According to the first one, i) the polyions were represented as
freely jointed chains of charged hard spheres. The
counterions were modeled as charged hard spheres and the solvent
as a continuum with a dielectric constant of pure water under
conditions of observation. The second model ii) differed from i) only
by the description of the oligoions; the beads of each oligoion were
ordered in a straight line and remained totally rigid during the
simulation. Computer simulations were performed for
three different degrees of polymerization; $N_{\mathrm{m}}$ (number of beads)
was 16, 32, and 48. In order to test the effect of the size of the
counterion (hydration-dehydration effect), we also varied its radius
$r_{\mathrm{c}}$. The two experimentally most thoroughly studied thermodynamic properties, i.e., the
enthalpy of dilution (see, for example,~\cite{vesnaver,cebasek} and the
activity coefficient~\cite{wandrey}, were examined as functions of
polyelectrolyte concentration. In addition, the pair correlation
functions between the charged sites and counterions in solution, as
well as the sites and ions themselves, were monitored and used to
explain the calculated properties. The simulation results are discussed
in view of the experimental results for these systems.

\section{Models and simulations}

In the present work, two different models were used. In the first
model, flexible charged oligomers are made of tangentially bonded charged
hard spheres of radius 2~{\AA}, while in the second model the monomer
units (beads) of the same size are ordered in a straight line and remain totally
rigid (see figure~1~(b)) during the simulation. Counterions are in both
models represented as charged hard spheres. Each monomer unit carries
one negative elementary charge, while monovalent counterions are positively
charged.  The total number of counterions is thus equal to the total number of
monomer units due to electroneutrality.
We systematically varied i) the oligomer size, which was 16, 32, and 48 monomer
units for both models, ii) the counterion's size, which was 1, 2, and 3~{\AA},
and iii) the monomer concentration from $5 \cdot 10^{-5}$ to 0.2 mol/L.

Interaction between the charged sites is governed by Coulomb
and hard sphere potentials:
\begin{equation}
u_{ij}(r_{ij})= \left\{ \begin{array}{cc}
{\frac{Z_iZ_j L_{\mathrm{B}}} %\over
{r_{ij}}}\,, & r_{ij}\geqslant {R_i+R_j}\,, \\
\infty, & r_{ij}<{{R_i+R_j}}\,.
\end{array} \right .
\end{equation}

Since the beads, as well as the counterions, carry one elementary charge,
Z$_i$ and Z$_j$ can be either  +1 (for the counterion) or
$-$1 (monomer unit). Note that the model used here is
essentially the same as studied by Chang and Yehtiraj~\cite{osm} and later by Bizjak et al.~\cite{bizjak}. The model
examined by Stevens and Kremer~\cite{kremer2,kremer3} is
slightly different but the two models provide essentially the
same insights.

The computer simulations were performed using the same code as
before~\cite{molsim,bizjak}. Simulations were carried out in a cubic box,
whose size was bigger than the size of oligomer. Long range Coulomb
potential was treated using the Ewald summation technique. The number
of particles in the simulation box was from 1280 to 5760,
depending on the chain lengths and concentration of monomers. For
flexible polyions we used three types of moves. Chains were ``moved''
by using a) translation (the whole chain was translated for particular
random vector), b) rotation of the shorter end of the chain and c)
bend stretching.  For rigid chains, the step a) and rotation of the
chain around its center were utilized.  The counterions were moved
through the box using simple translation procedure. The equilibration
run required $1\cdot 10^5$, and to
obtain good statistics for the production run, $1\cdot
10^6$ moves per particle were needed.

\section{Results and discussion}

Numerical results in this section apply to the concentration of
monomer units \linebreak $c_{\mathrm{m}}=0.001$~monomol/dm$^3$. Further, $T = 298.15$~K, where $L_{\mathrm{B}}$ is equal to ${}= 7.14$~\AA , which yields the linear
charge density $\lambda = 1.785$.  In this way, $\lambda> 1$, this
is considered to be a moderately strong polyelectrolyte for which the
Manning condensation~\cite{manning0,manning1} should take place. To
orient the reader, $\lambda$ for poly(styrene)sulfonic acid is 2.8, and
for the DNA it is 4.2, while for another class of frequently studied
polyelectrolytes, i.e. $x,y$-ionenes~\cite{nagaya,Luksic1,Luksic2,Luksic3},  it is from  0.673 to
1.437. Only fully charged oligomers~-- where each bead carries one
negative elementary charge~-- were studied here. The effect of the
polyion length was explored with varying number of segments $N_{\mathrm{m}}$ from
16 to 48.

\subsection{Pair distribution functions}

Here we are interested in how an increase in the flexibility of polyions
and the size of counterions affect the spatial distributions of particles
in such systems. In the binary systems like these ones we have three
pair distribution functions (pdf): the monomer-monomer
$g_{\mathrm{mm}}(r)$, monomer-counterion $g_{\mathrm{mc}}(r)$, and
counterion-counterion $g_{\mathrm{cc}}(r)$ function. These correlation
functions have been calculated for both models and for two (or in
some cases three) values of the counterions radius, $r_{\mathrm{c}}$. Only the
most interesting pair distribution functions are presented here.

\subsubsection{Flexible oligoions}

In figure~2~(a)--(c) we present the results for the intermolecular part of the
monomer-monomer distribution function for $N_{\mathrm{m}} = 16$, 32, and
48, all for two different values $r_{\mathrm{c}}$ (1 and 3~{\AA}), as
indicated in the caption. This result shows that smaller
counterions (1~{\AA}), due to their stronger electrostatic
screening of polyions, cause the monomer units belonging to
different oligoions to distribute at distances
(continuous curve)  smaller than larger ions 3~{\AA} (dashed curve).
The full monomer-monomer distribution function, which also contains
 the intramolecular part, is shown in figure~3~(a)--(b).
\begin{figure}[!h]
\begin{center}
\includegraphics[clip=true,width=7cm]{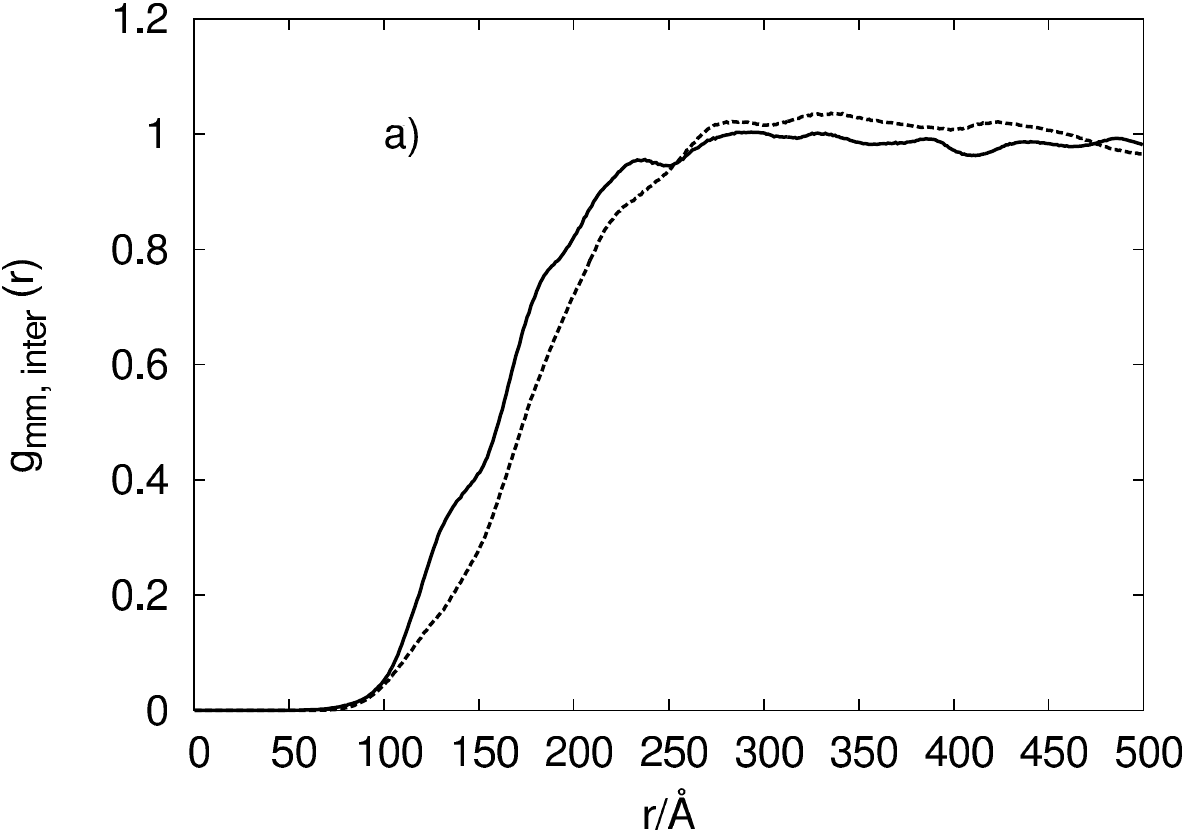}
\includegraphics[clip=true,width=7cm]{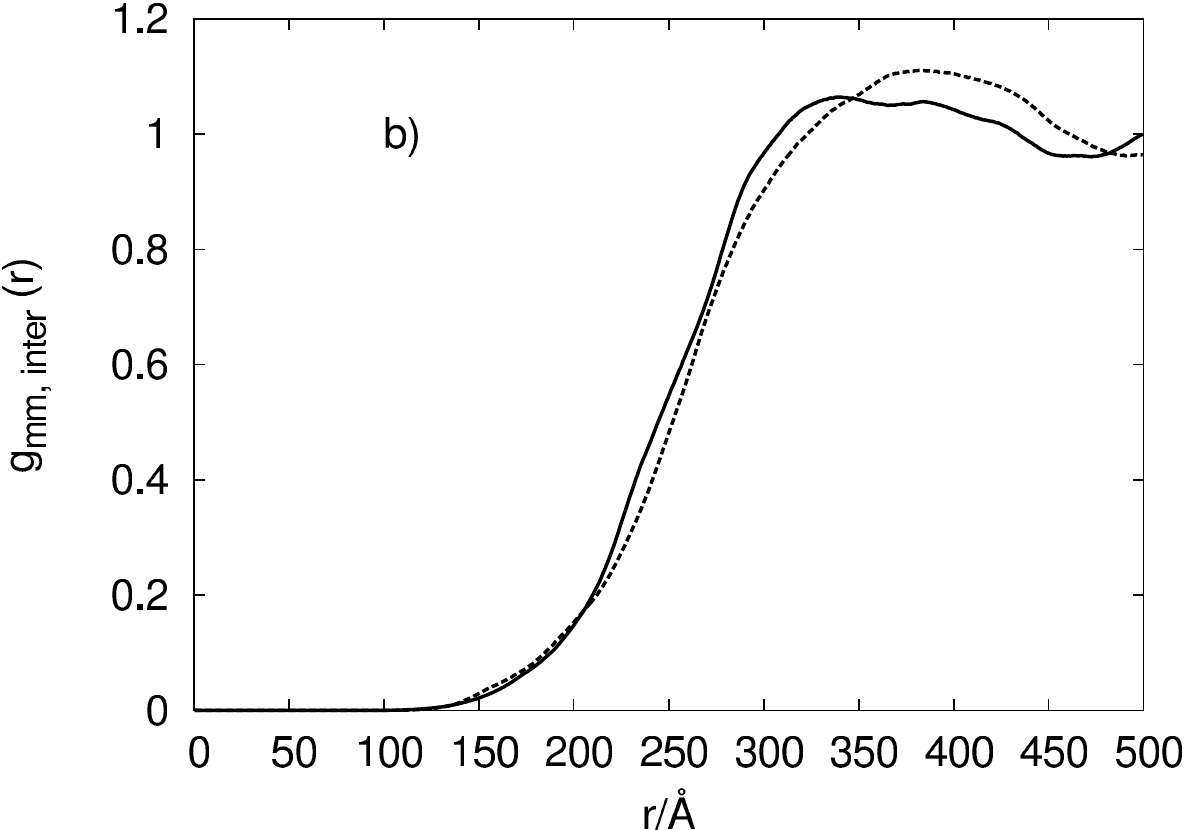}
\includegraphics[clip=true,width=7cm]{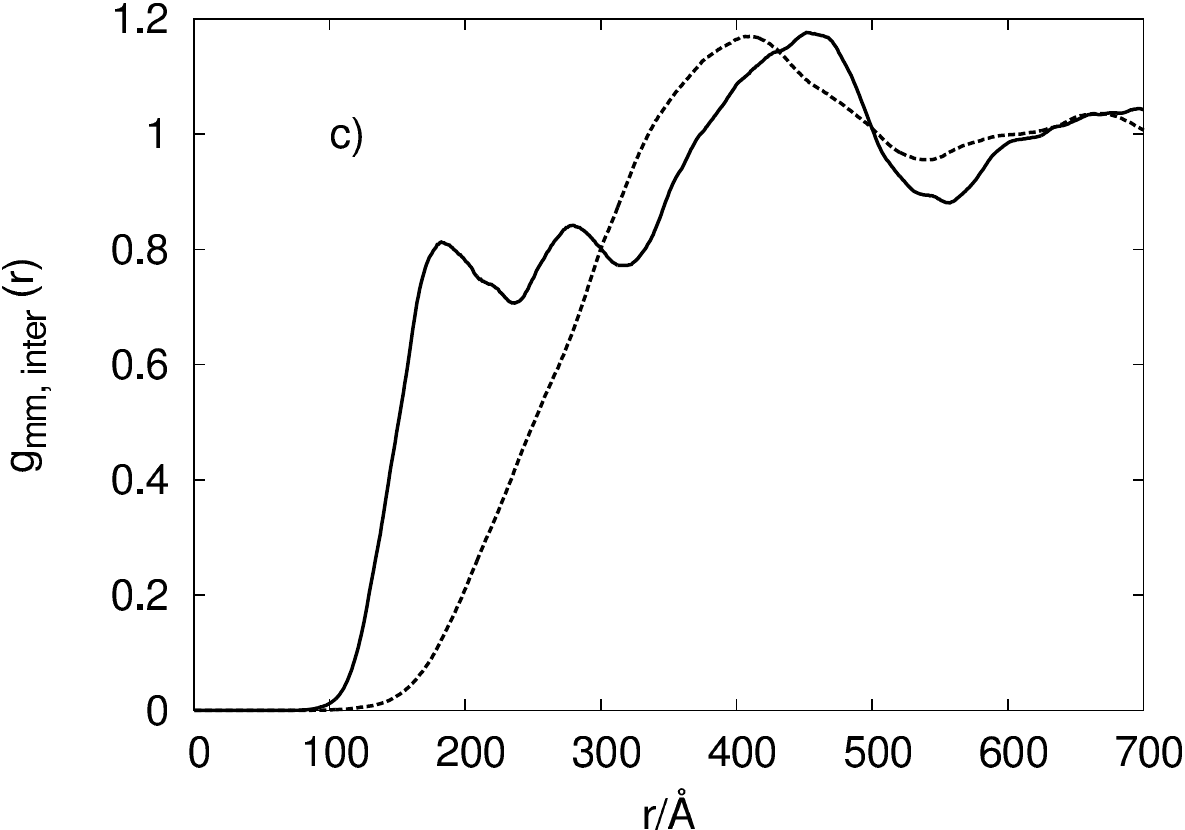}
\caption{Intermolecular monomer-monomer pair distribution function at N = 16 (a), 32 (b), and 48 (c),
for $r_{\mathrm{c}}$ = 1~\AA~(continuous curve) and 3~\AA~(dashed curve). }
\label{fig:m-m}
\end{center}
\end{figure}
\begin{figure}[!h]
\includegraphics[width=0.48\textwidth]{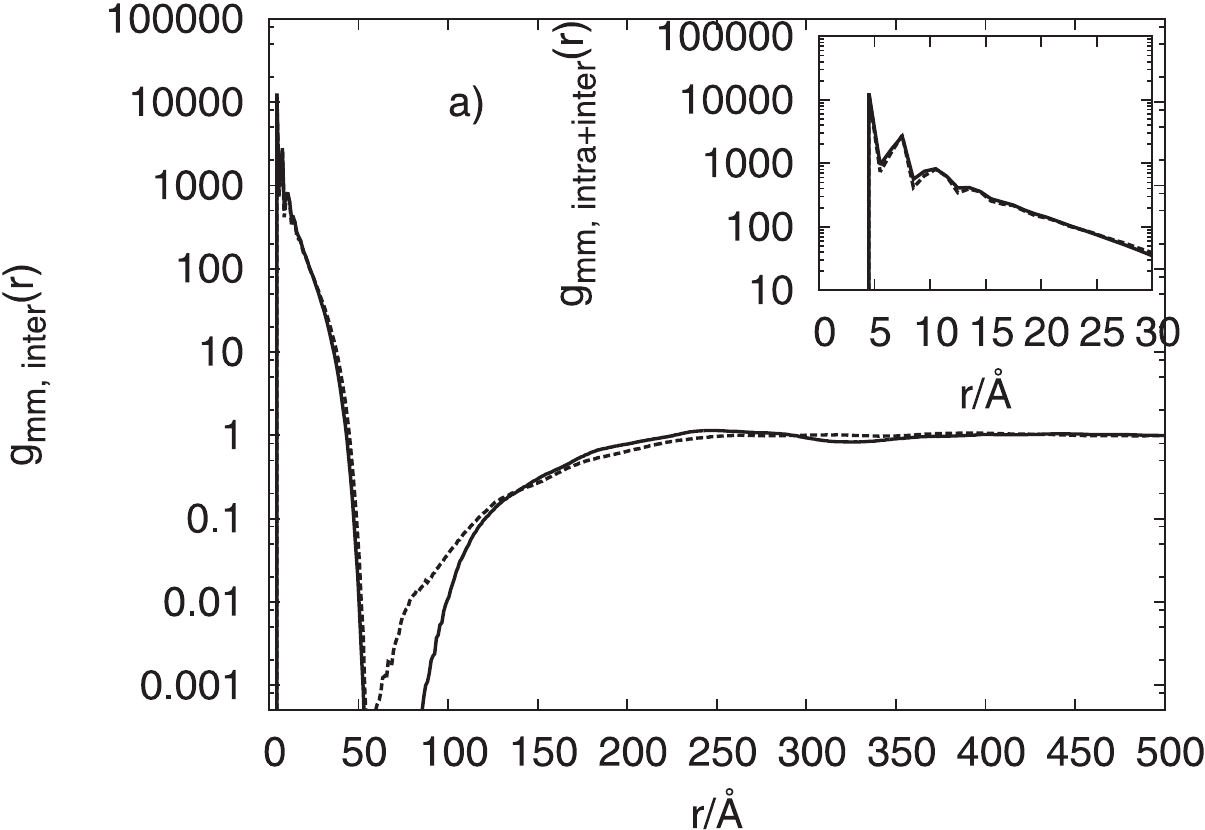}%
\includegraphics[width=0.48\textwidth]{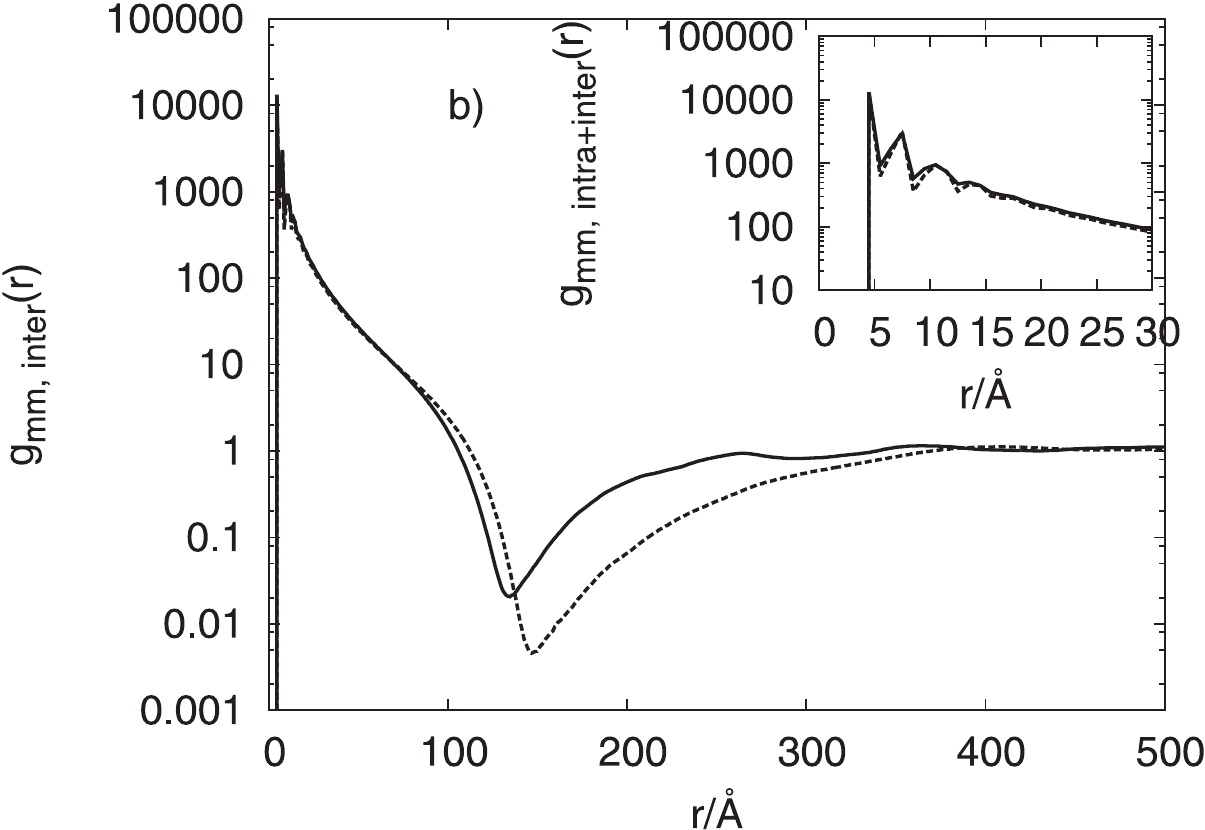}%
\caption{Intermolecular monomer-monomer pair distribution function for $N_{\mathrm{m}} = 16$
(a) and $N_{\mathrm{m}} = 48$ (b), for $r_{\mathrm{c}} = 1$~\AA (continuous line) and 3~\AA (dashed line). Insets
show the total (intermolecular plus intramolecular) distribution at short separations.}
\label{fig:m-m2}
\end{figure}
\begin{figure}[!h]
%\vspace{-5mm}
\begin{center}
\includegraphics[clip=true,width=0.48\textwidth]{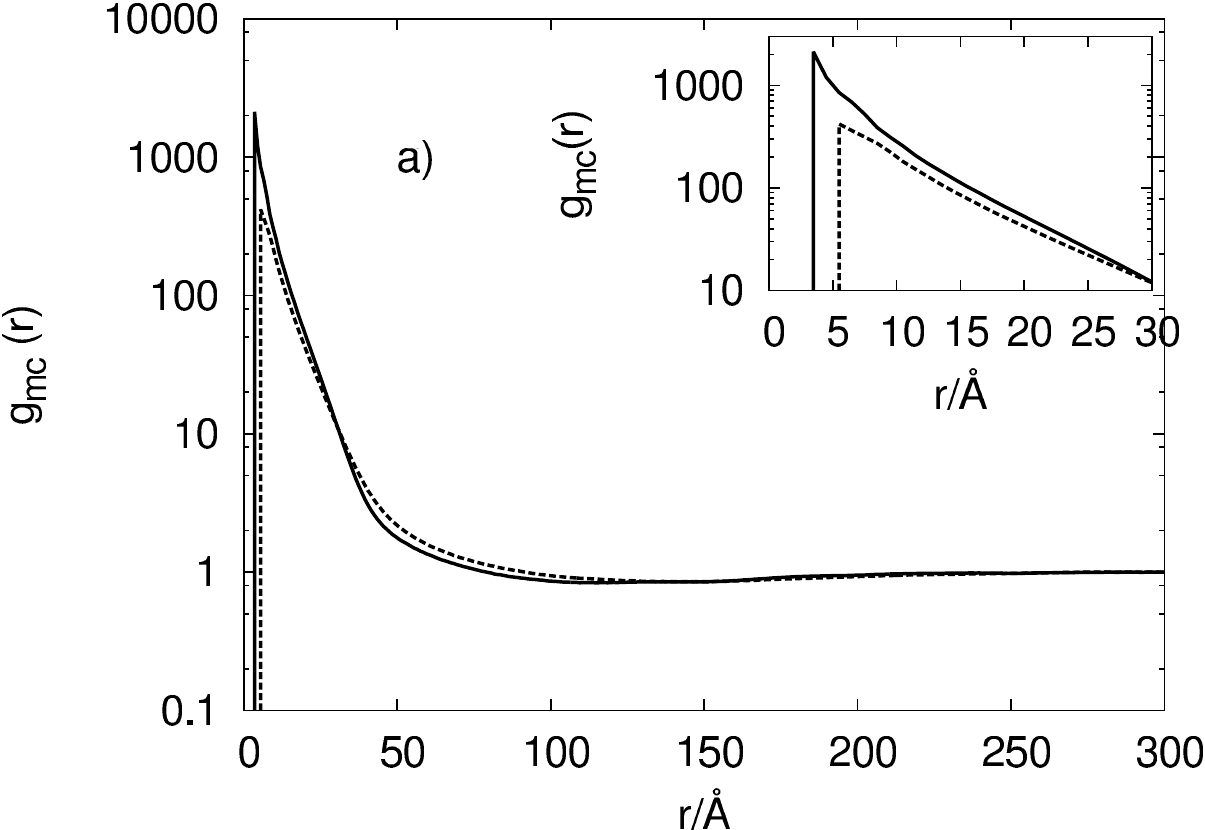}
\includegraphics[clip=true,width=0.48\textwidth]{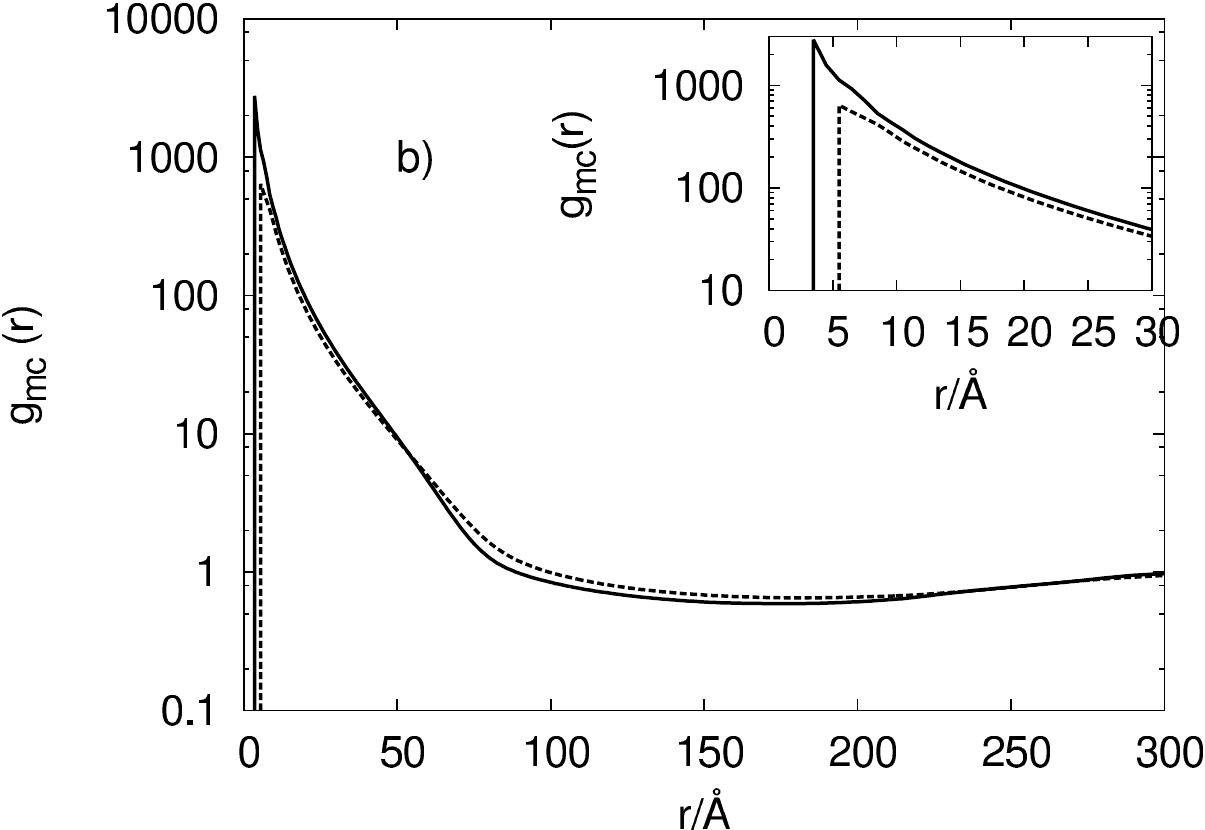}
\includegraphics[clip=true,width=0.48\textwidth]{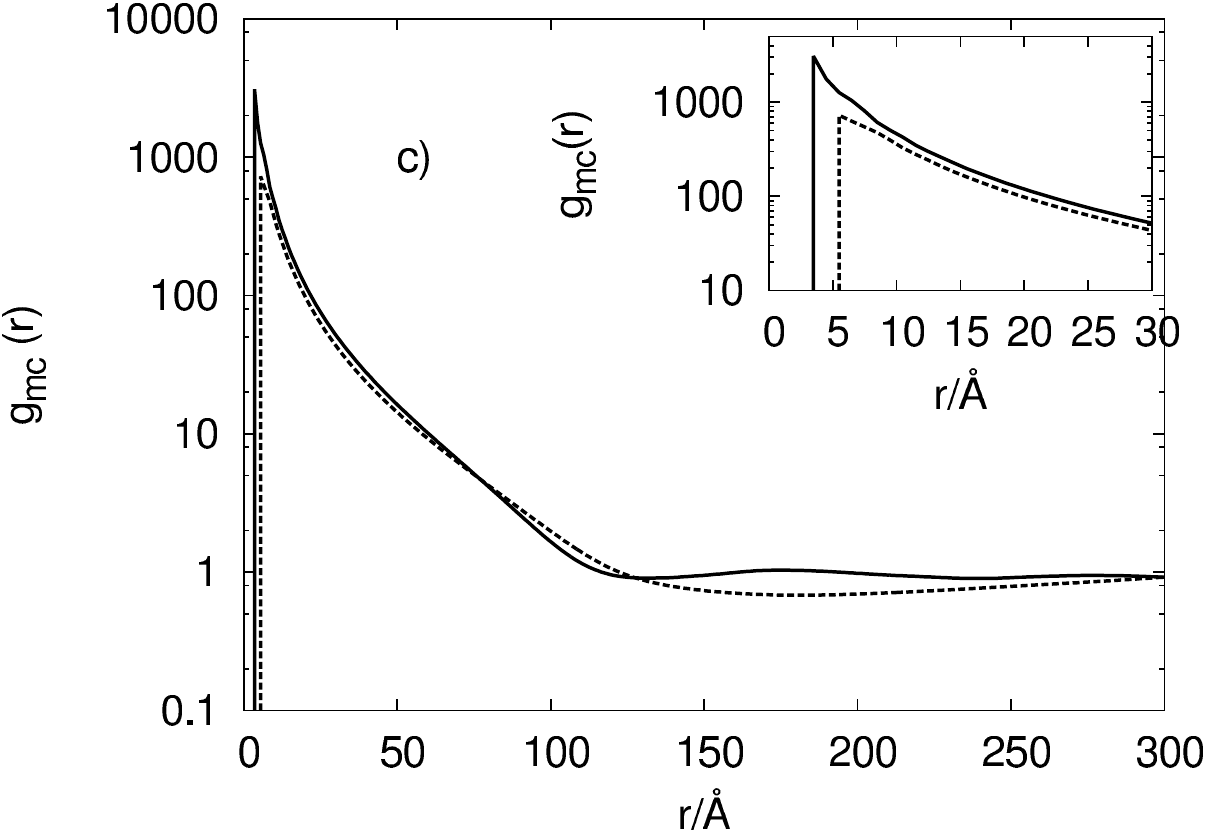}
\caption{Monomer-counterion pair distribution function at $N_{\mathrm{m}} = 16$ (a) , 32 (b) , and 48 (c),
all for $r_{\mathrm{c}} = 1$~\AA (continuous line) and 3~\AA (dashed line). }
\label{fig:m-c}
\end{center}
\end{figure}
\begin{figure}[!h]
\vspace{-2mm}
\begin{center}
\includegraphics[clip=true,width=0.48\textwidth]{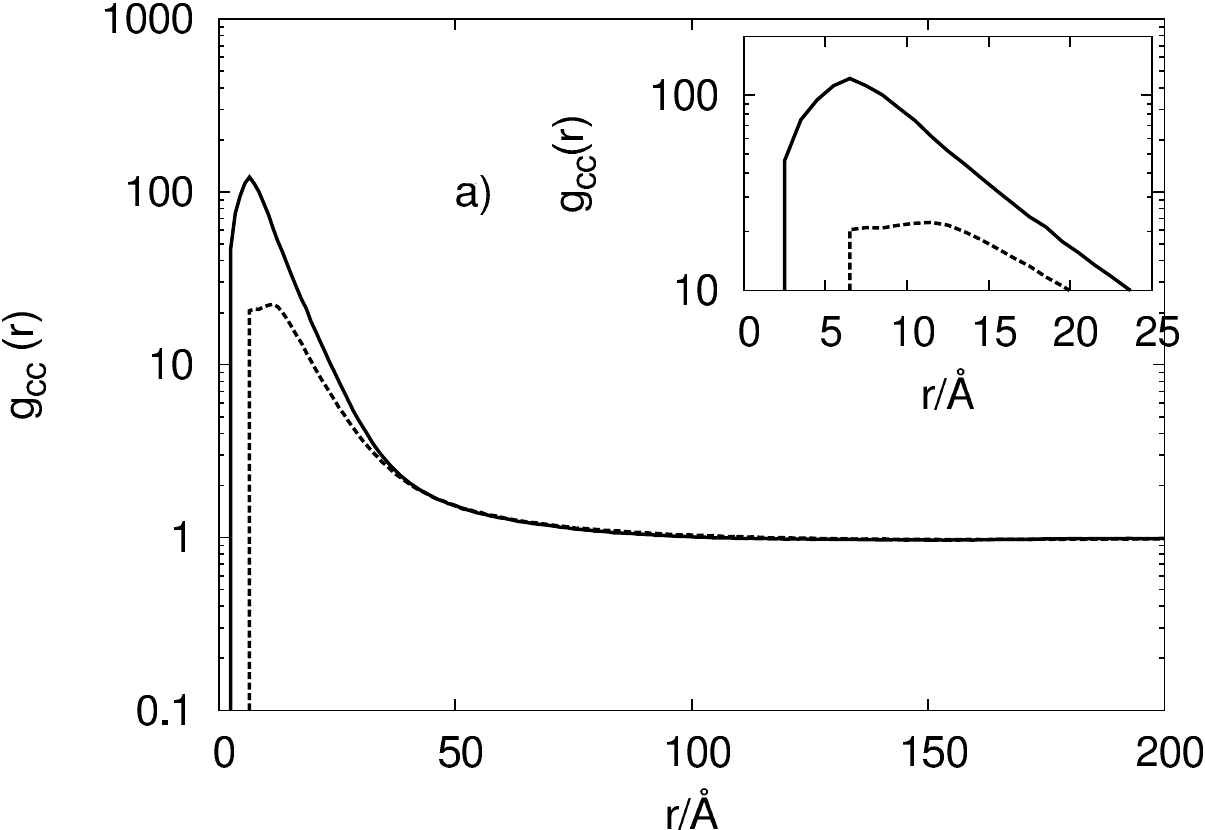}
\includegraphics[clip=true,width=0.48\textwidth]{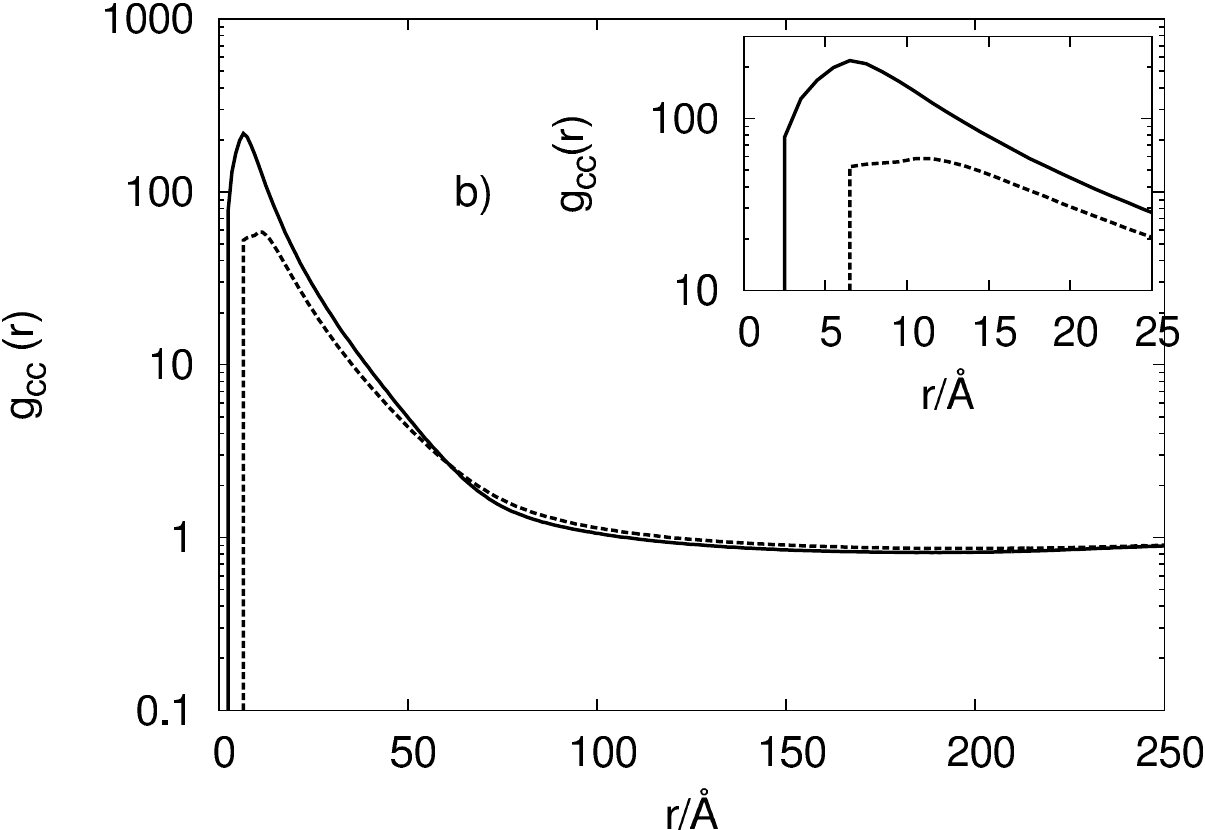}
\includegraphics[clip=true,width=0.48\textwidth]{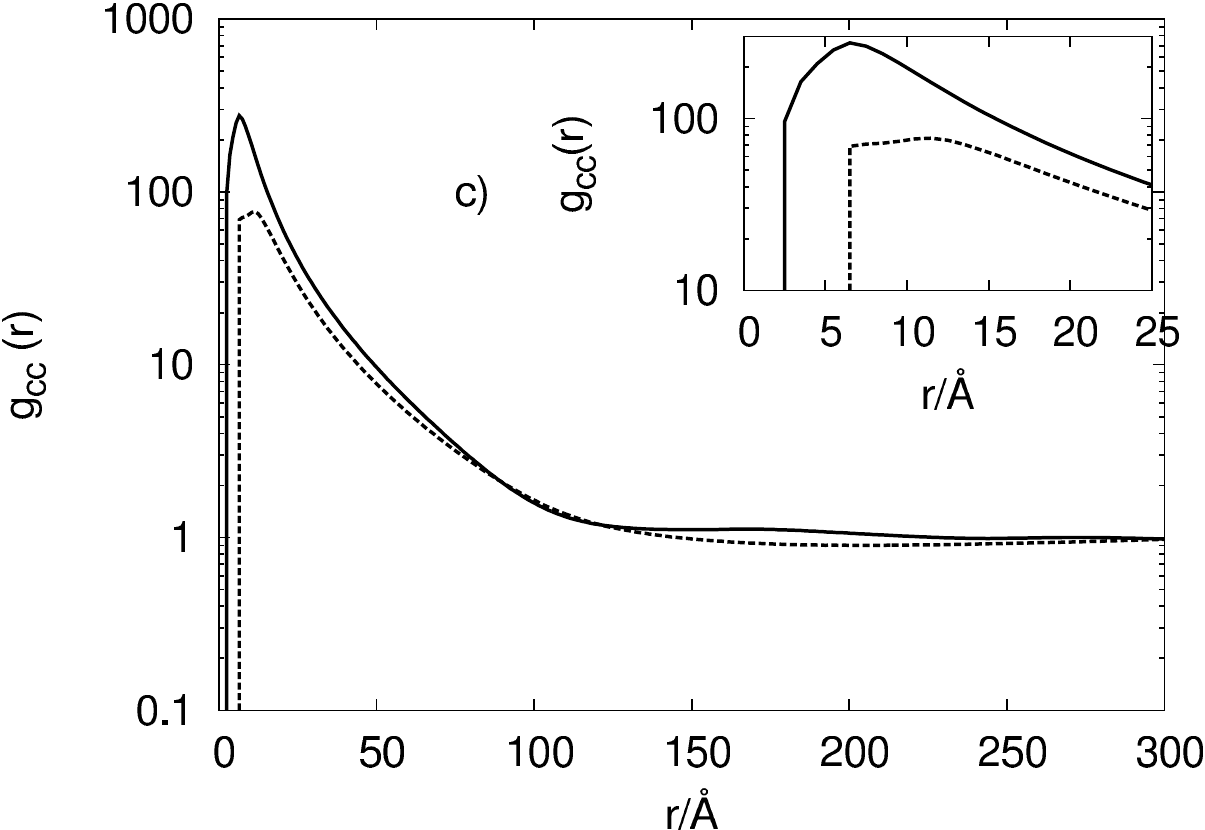}
\caption{Counterion-counterion pair distribution function at  $N_{\mathrm{m}} = 16$ (a), 32 (b),
and 48 (c) all for $r_{\mathrm{c}} = 1$~\AA (continuous line)  and 3~\AA (dashed line). }
\label{fig:c-c}
\end{center}
\vspace{-13mm}
\end{figure}

The monomer-counterion pair distribution functions, shown in figure~4~(a)--(b), reflect strong counterion-oligoion attraction. These results
apply to $N_{\mathrm{m}} = 48$.  The counterions having smaller radius are capable of closer approaching the oligoion, and thus the peak of the pair distribution
function reaches higher values.  The minimum of $g_{\mathrm{mc}}(r)$ at large
distances, i.e. at $r$ $\approx$ 200~{\AA} is an interesting new
feature still unknown within the Poisson-Boltzmann cell model
calculations. The minimum was  also found for $N_{\mathrm{m}} = 16$ (and 32)
but it was shallower and located closer to the origin.

The third pair distribution function, shown in figure~5~(a)--(c),
describes the counter\-ion-counter\-ion correlation, $g_{\mathrm{cc}}(r)$.
In several
recent papers~\cite{jesus1,jesus2,jesus2a,jesus3}, this distribution function has been  extensively studied for the cylindrical model
with infinitely long polyion. A strong
increase of concentration of equally charged counterions in
the vicinity of the polyion forms a basis of the catalytic
effect~\cite{morawetz1,morawetz2}. The shape reflects the fact
that the highest probability of finding two counterions is just
on the opposite sides of the charged bead. Such distributions
of counterions are not limited to linear polyelectrolyte and
may  also appear in highly charged micellar solutions~\cite{RVH}.

\subsubsection{Rigid versus flexible oligoions}

First studies of short cylindrical polyions, with the purpose of exploring the effect of an increased length of the oligoion on the osmotic
coefficient, were published long ago~\cite{VD,BD}. The Monte Carlo
method and the Poisson-Boltzmann theory were used for this
purpose. These calculations were performed in the cell model
approximation and the first simulation of an isotropic solution of
rigid-chain polyelectrolyte seems to be due to Antypov and Holm~\cite{antypov}.  These authors, however, did not present any spatial
distribution functions; the pair correlation functions for our system are
shown next. In particular, we are interested in the comparison of the
correlation functions obtained for flexible and rigid oligoions, which is
presented next.

First in figure~6~(a)--(c) we display the result for the interparticle part of the
monomer-monomer pair distribution function as obtained by the
flexible and rigid oligoion models. The results show that monomers of
flexible oligoions are generally distributed at  distances larger than
those of rigid oligoions.
\begin{figure}[ht]
\begin{center}
\includegraphics[clip=true,width=0.48\textwidth]{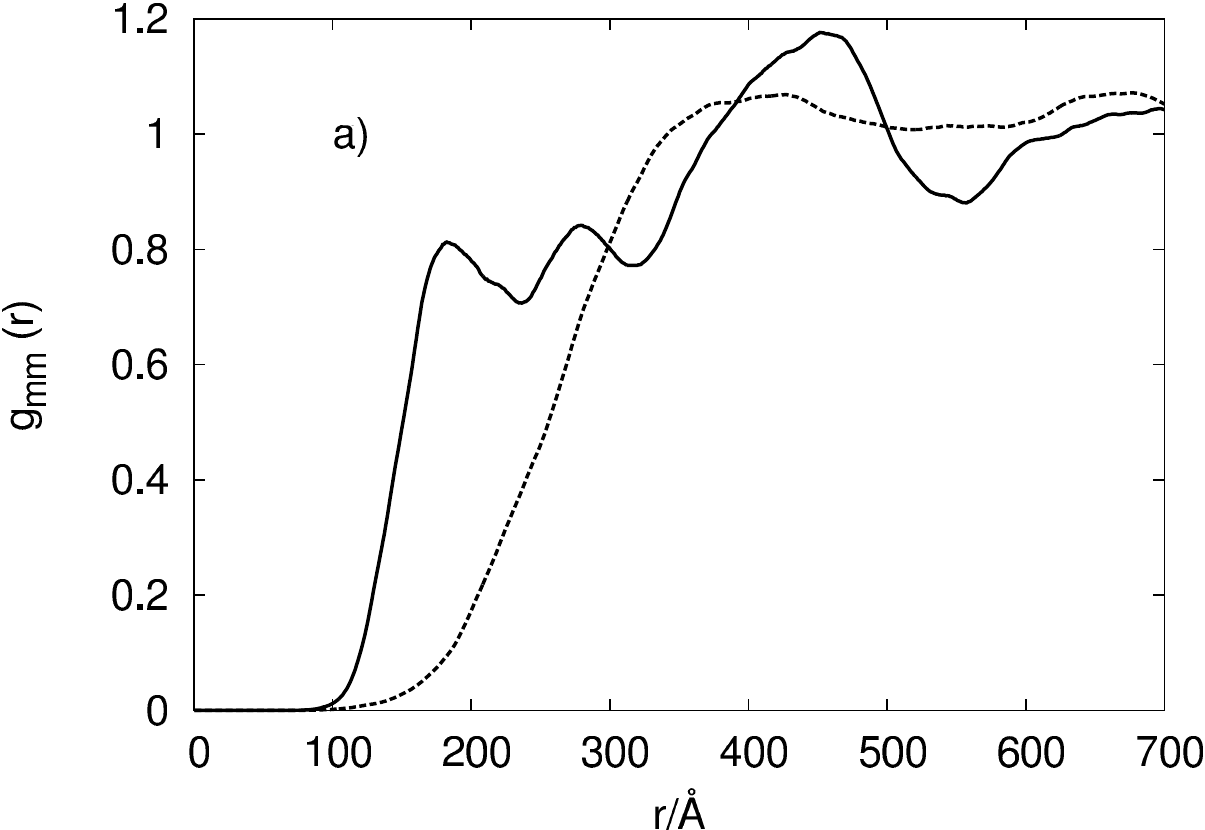}
\includegraphics[clip=true,width=0.48\textwidth]{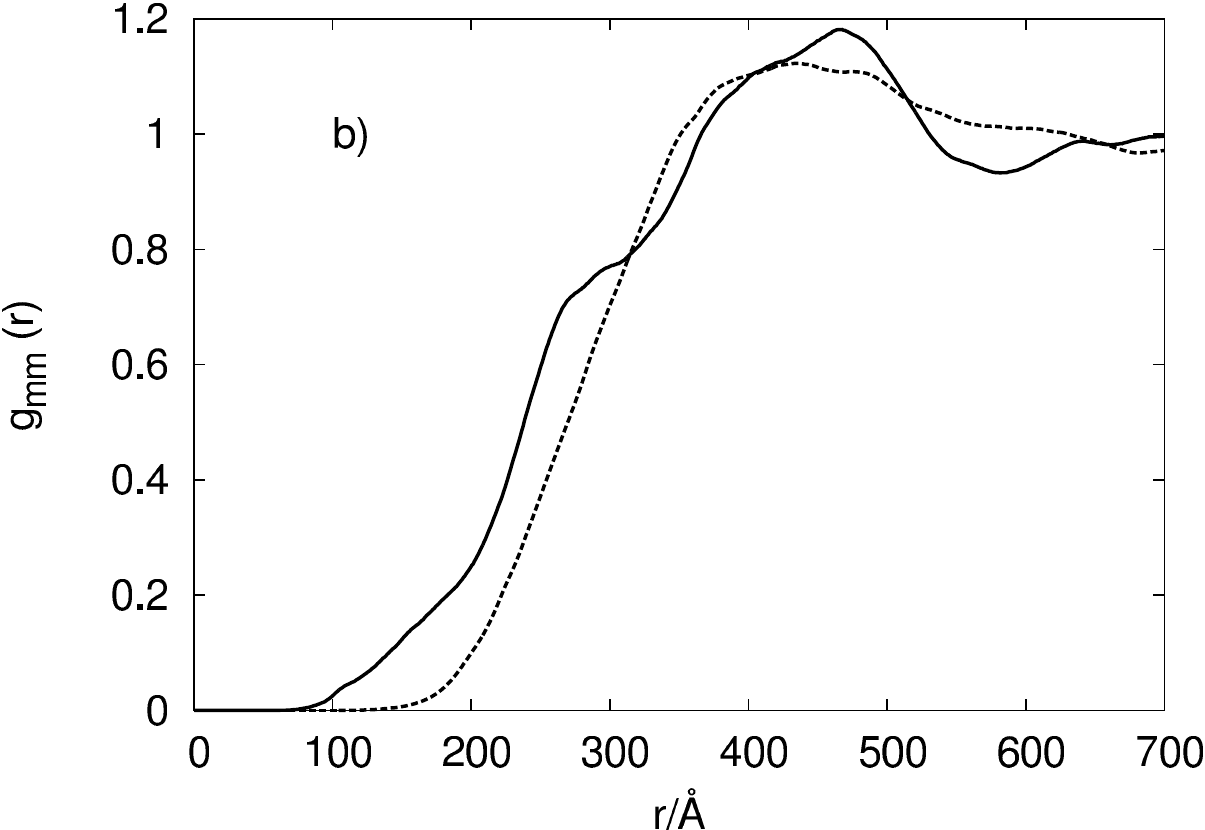}
\includegraphics[clip=true,width=0.48\textwidth]{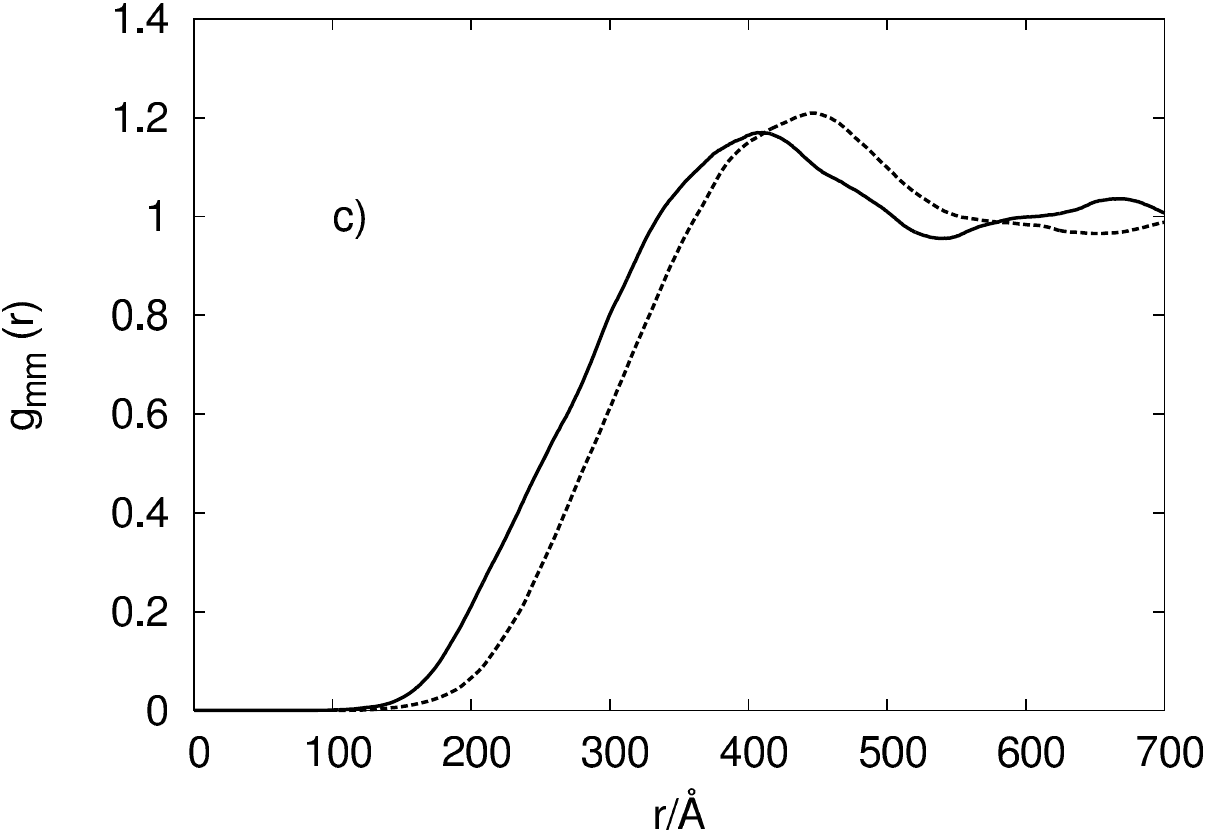}
\caption{Intermolecular monomer-monomer pair distribution function for $N_{\mathrm{m}} = 48$ at $r_{\mathrm{c}} = 1$~\AA~(a), 2~\AA~(b),
and 3~\AA~(c) for flexible (continuous  line) and rigid chains (dashed line).}
\label{fig:mm48}
\end{center}
\end{figure}
The other distribution functions are shown in figures~7 and 8. First, the
monomer-counterion distributions are displayed in figure~7. As we see
the distributions differ qualitatively; the descent of the function
obtained for rigid polyions is steeper at the beginning and forms a
much shallower minimum.  At a certain distance, the two functions
intersect each other. Up to this distance, the concentration of
counterions near the rigid polyion is significantly higher. Different
shapes of the pair distributions, of course, affect the fraction of
counterions calculated within a certain distance from the polyion.

\begin{figure}[!h]
\begin{center}
\includegraphics[clip=true,width=0.48\textwidth]{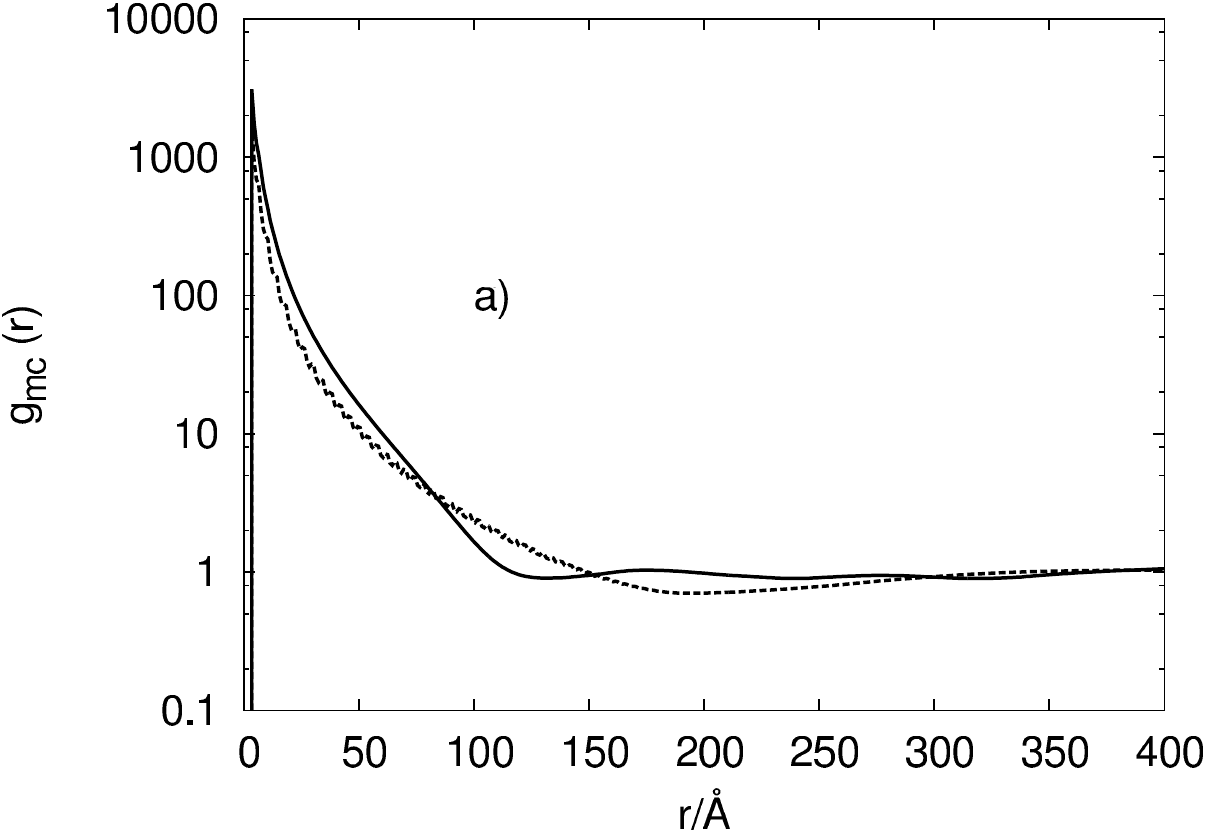}
\includegraphics[clip=true,width=0.48\textwidth]{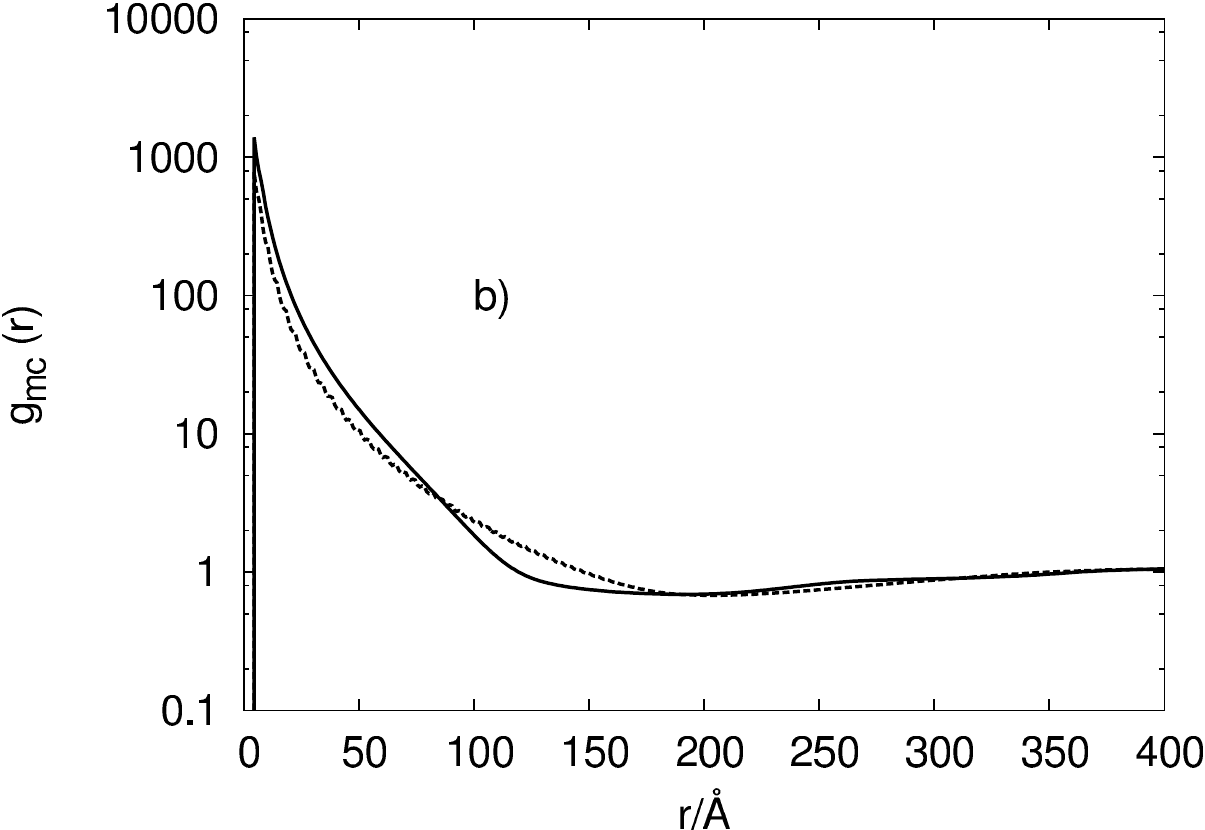}
\includegraphics[clip=true,width=0.48\textwidth]{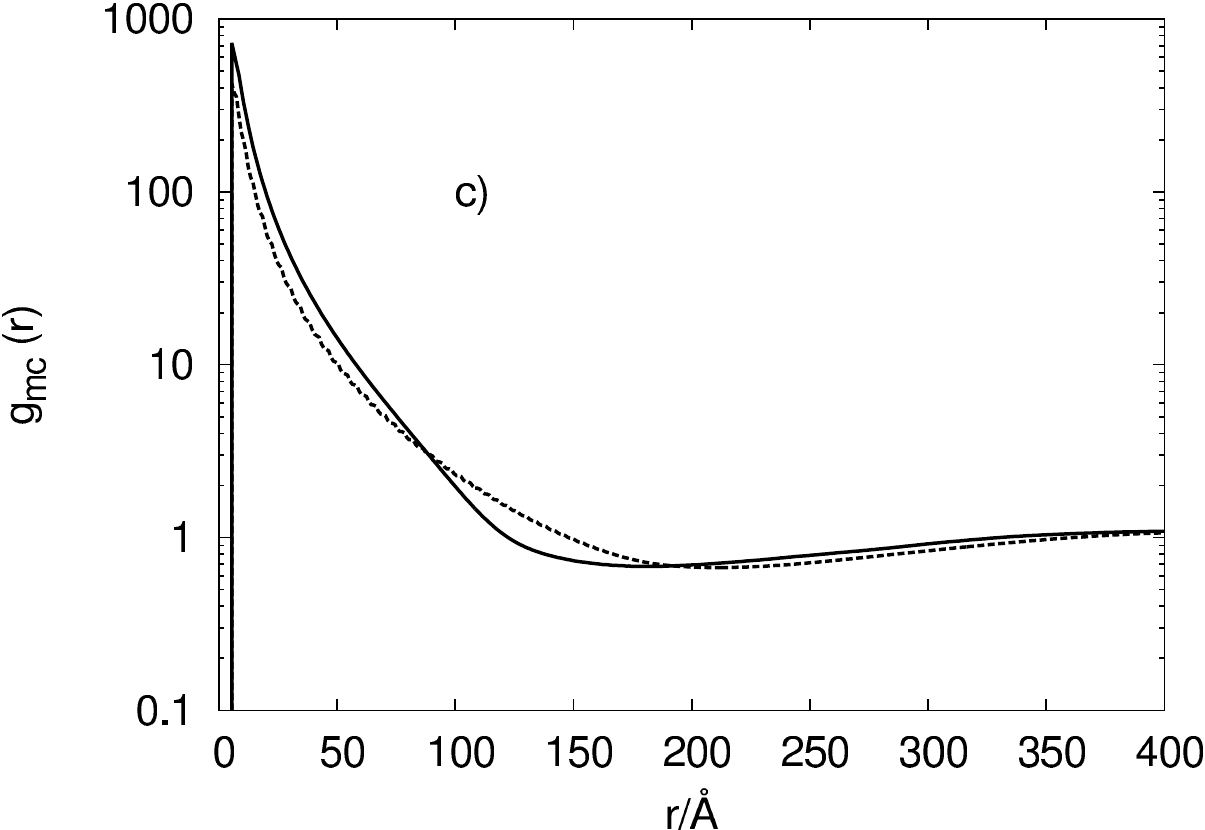}
\caption{Monomer-counterion pair distribution function  for  $N_{\mathrm{m}} = 48$ at  $r_{\mathrm{c}} = 1$~\AA~(a),
2~\AA~(b), and 3~\AA~(c) for flexible (continuous line) and rigid chains (dashed line).}
\label{fig:mc48}
\end{center}
\end{figure}
\begin{figure}[!h]
\vspace{-2mm}
\begin{center}
\includegraphics[clip=true,width=0.48\textwidth]{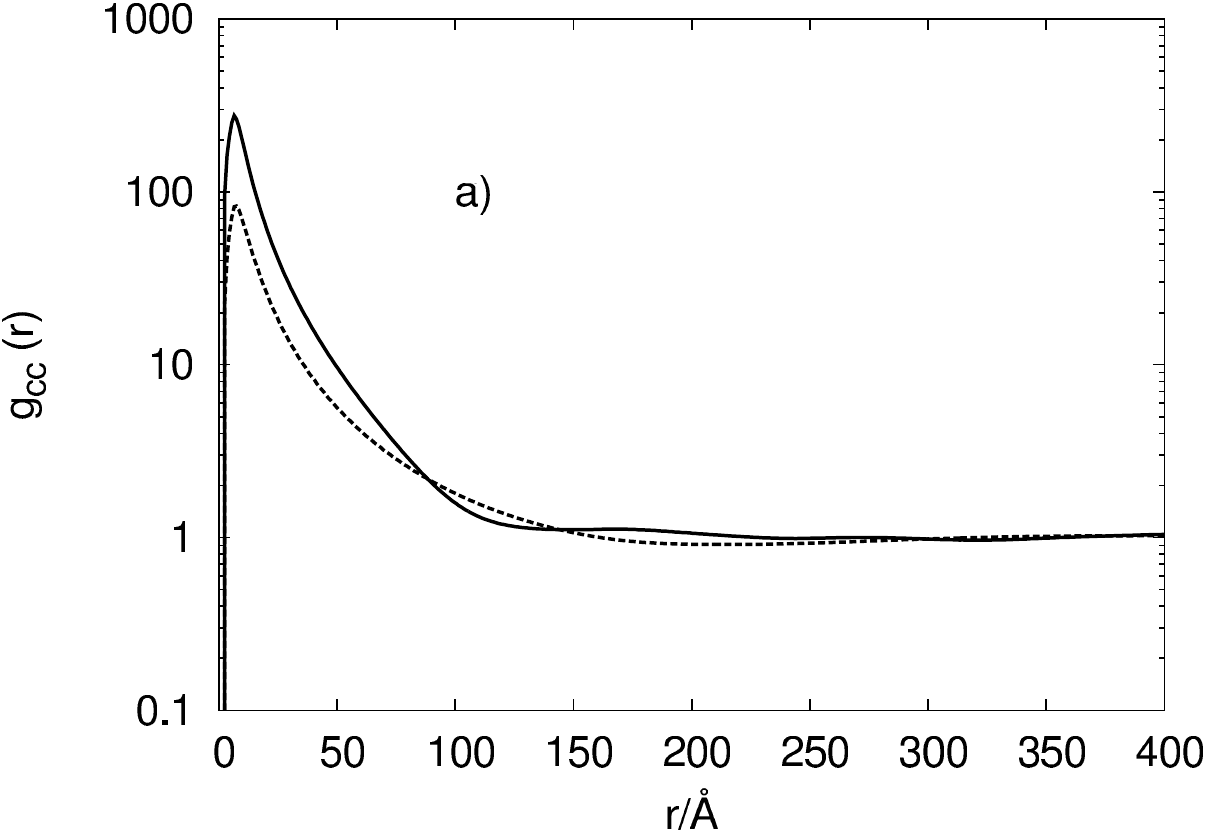}
\includegraphics[clip=true,width=0.48\textwidth]{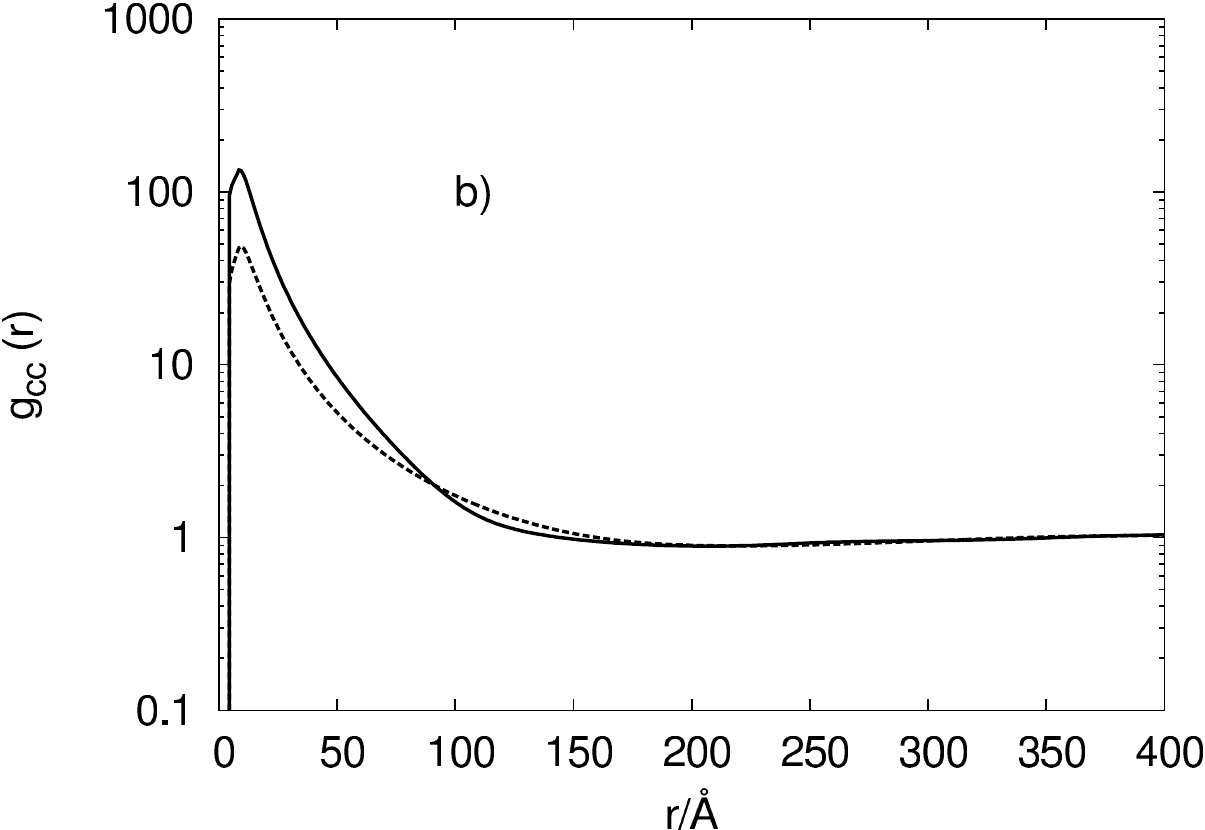}
\includegraphics[clip=true,width=0.48\textwidth]{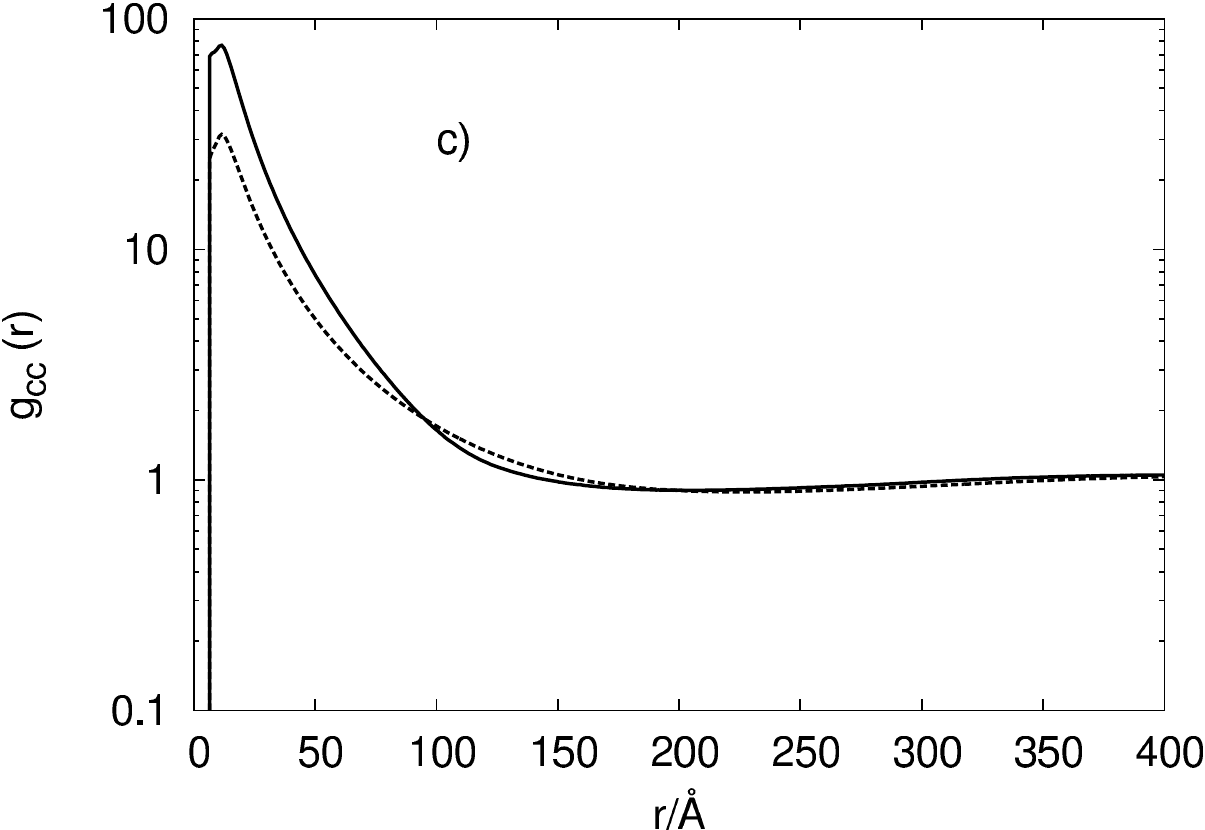}
\caption{Counterion-counterion pdf for $N_{\mathrm{m}} = 48$ at  $r_{\mathrm{c}}  = 1$~\AA~(a), 2~\AA~(b), and 3~\AA~(c)
for flexible (continuous line) and rigid chains (dashed line).}
\label{fig:cc48}
\vspace{-13mm}
\end{center}
\end{figure}

In figure~8 we show the counterion-counterion, $g_{\mathrm{cc}}(r)$, distribution
functions for both cases. Similarly, we observe significant
differences between the functions obtained for rigid and flexible
polyions. The correlation function for rigid polyions (dashed line) runs
below the one for flexible polyions (continuous line), indicating weaker
correlation between equally charged ions in the vicinity of the polyion.
This observation is only true to a certain distance where, as shown
 in figure~7, the two distribution functions intersect. As found in
previous studies of infinitely long and rigid polyions~\cite{jesus2},
smaller ionic size yields stronger correlation between the counterions
in the electrical double-layer. This counterion-counterion correlation
is neglected in the Poisson-Boltzmann theory  but as shown before~\cite{jesus1} (see also the references therein), this is not considered
to be a serious source of error when  monovalent counterions are
present in solution. For divalent counterions this is not true any more.
In such cases, the counterion-counterion correlation may be strong enough to cause clustering of the polyions; this effect cannot be
accounted for by the mean-field Poisson-Boltzmann theory.

\subsection{Conformational averages}

The quantities, reflecting the average conformation of the oligoion in
solution, are the end-to-end distance and the average radius of
gyration. Both properties have been for flexible oligoions studied
before~\cite{bizjak}, but only for one size of the counterions. Here, the
study is extended to the counterions with $r_{\mathrm{c}}$ = 1 and 3~\AA{}
in order to quantify the effects of the counterion size on these
averages.

The average values of the end-to-end distance, $R_{\mathrm{ee}}$ (in \AA)
and the radius of gyration $R_{\mathrm{g}}$ (in \AA) are, as a function of the
counterion size and for the degrees of polymerization $N_{\mathrm{m}}$ equal to
16, 32, and 48, collected  in table~1.  For short oligoions ($N_{\mathrm{m}}=
16$)  the values of $R_{\mathrm{g}}$ and $R_{\mathrm{ee}}$ do not change much with the
change of the counterion size. Shorter oligoions tend to assume more
extended conformations in solution. Due to their low degree of
polymerization, they are surrounded by a smaller number of
counterions. The electrostatic forces between the oligoion and
counterions in solution are in such situations less strong, and the
variations of the counterion size only have small effects.
\begin{table}[ht]
\caption{Conformational averages and excess chemical potential of counerions at $c_{\mathrm{m}}=0.001$~mol/L
as a function of $N_{\mathrm{m}}$ and for three different $r_{\mathrm{c}}$ values for flexible chains.}\vspace{1ex}
\begin{center}\begin{tabular}{|c|c|c|c|c|c|c|c|c|c|}
\hline
\hline
$N_{\mathrm{m}}$   & \multicolumn{3}{|c|}{16}  & \multicolumn{3}{|c|}{32}  & \multicolumn{3}{|c|}{48}  \\
\hline
$r_{\mathrm{c}}$ [\AA]  &    1   &    2   &   3   &    1   &   2   &   3   &     1   &    2   &    3   \\
\hline
$R_{\mathrm{ee}}$ [\AA] &  36.64 &  38.18 & 39.27 &  72.74 & 76.30 & 79.12 &  106.94 & 113.76 & 118.48 \\
$R_{\mathrm{g}}$  [\AA] &  12.77 &  13.21 & 13.50 &  24.42 & 25.46 & 26.27 &   35.33 &  37.37 & 38.70  \\
$-\beta \mu^{\mathrm{ex}}$ & 0.479 & 0.410 & 0.339 & 0.693 & 0.613 & 0.542 & 0.891 & 0.793 & 0.707 \\
\hline
\end{tabular}
\end{center}\end{table}

An increase of the degree of polymerization increases $R_{\mathrm{ee}}$ and
$R_{\mathrm{g}}$. Also, larger counterions do not screen the repulsion
between the monomer units (beads) as well as the smaller ones;
the effect increases the end-to-end distance. It is of interest to
compare the end-to-end distance for a certain oligoion, with its
contour (fully extended)  length. For $N_{\mathrm{m}}=48$, the fully extended
oligoion measures 192.0~\AA, while the average end-to-end distance
is only around 118.5~\AA.

\subsection{Enthalpy of dilution}

For polyelectrolyte solutions the excess internal energy is not
that important {\it per se}, however, the difference of
$U^{\mathrm{ex}}$ between two concentrations, can be approximated with
the enthalpy of dilution, $\Delta_{\mathrm{d}} H$. This quantity can be
accurately measured~\cite{cebasek,lipar,Boyd,vesnaver,cation,pohar,vlachy2008},
providing important insights into the nature of interaction in
polyelectrolyte solutions. The measurements performed for
various polyelectrolyte systems indicated important deviations
from theoretical predictions (for review see,~\cite{vlachy2008}). For example, the classical electrostatic
theories predict exothermic effects upon dilution~\cite{dolar},
while experimental data show that $\Delta_{\mathrm{d}} H$ can be negative or
positive depending on the nature of the counterion~\cite{cebasek,pohar}.

The results for $\Delta_{\mathrm{d}} H^{\mathrm{ex}}$ are
presented in figure~9 for different $N_{\mathrm{m}}$ as a function of the
${}-\log_{10}c_{\mathrm{m}}$. The results for both flexible and
totally rigid model of oligoions are presented in the same
figure. The most important observations are as follows:
 i)
the enthalpies are, as found in our previous study~\cite{bizjak}, negative and the slope depends on the degree of
polymerization; more heat is released upon dilution per mole of
monomer units for shorter polyions, ii) the $\Delta_{\mathrm{d}} H$
values are less exothermic for flexible polyions than for the
rigid ones, iii) the effects of counterion radius are
noticeable for solution of rigid polyions but much less
pronounced in case of flexible polyions. In all cases the
differences are small and, as our experience shows, they are within the
uncertainties of experimental data.
\begin{figure}[ht]
\begin{center}
\includegraphics[clip=true,width=9cm]{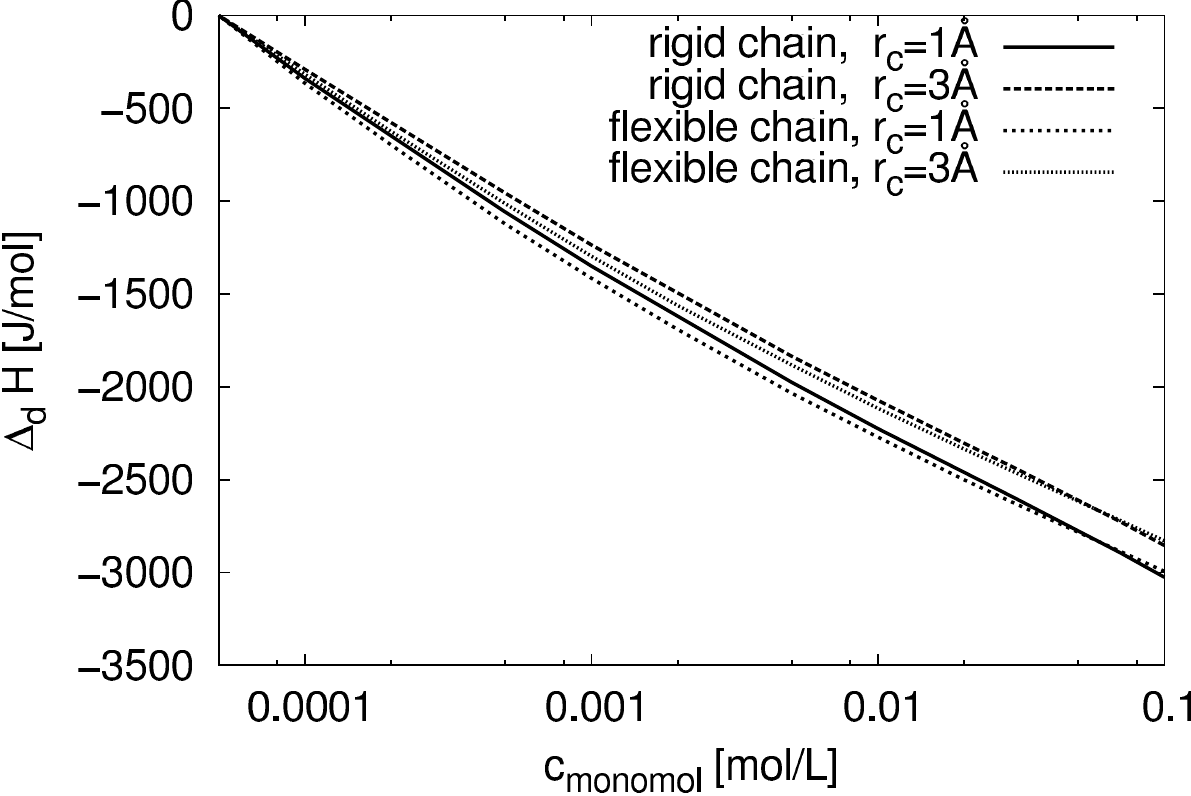}
\caption{Enthalpy of dilution as a function of $c_{\mathrm{m}}$ for $N = 32$ at $r_{\mathrm{c}} = 1$ and 3~{\AA}
for both flexible and rigid chains.}
\label{dil-H}
\end{center}
\end{figure}

We can compare these results with those presented, for example,
in figure~5 by Lipar et al.~\cite{lipar}. These authors have
measured the heats of dilution of various alkaline salts of
polyanetholesulphonic acid at 298~K.  As we see both the trends
and magnitude of the effect are correctly predicted by our
models.

\subsection{Activity coefficient of counterions}

A comprehensive examination of the counterion activity in
polyelectrolyte solutions was published by Wandrey et al.~\cite{wandrey}. The study revealed the effect of the degree of
polymerization on the counterion activity, the factor which was
not evident from the theories treating polyions as infinitely
long and rigid cylinders. Here we explore the effect of another
 factor of this kind, i.e., the polyion flexibility.  Excess chemical
potential (activity coefficient) of counterions $\mu^{\mathrm{ex}}$ has
been studied in our previous work~\cite{bizjak}. The novelty of
the present work lies in the simulations that we performed for both
i) flexible and ii) rigid models as well as in examining the effect
of the size of the counterions.

As before, we use Widom's method to determine this quantity. In
equation~\eqref{eq:widom}, $\Delta U$ represents the sum of pairwise
potentials of the inserted fictitious counterion with all others
counterions and monomer units (see, for example,~\cite{frenkel})
\begin{equation}
  \label{eq:widom}
  \beta \mu^{\mathrm{ex}}=- \ln\left\langle \rm{e}^{-{\Delta U}/{k_{\mathrm{B}} T}}\right\rangle.
\end{equation}

Although the individual activity coefficient of counterion  is, strictly speaking,
not a well-defined thermodynamic property, still it is of
theoretical interest. For the salt-free solutions in the cylindrical cell
model approximation, this quantity is equal to the osmotic coefficient~\cite{gueron}. Earlier experimental determinations of counterion
activity coefficient are due to Nagasawa and Kagawa~\cite{nagasawa}, Oman and Dolar~\cite{dolar,oman}, and Joshi and
Kwak~\cite{joshi}. More recent experimental results were, for solutions
of aliphatic ionenes with variable degree of charging, obtained by
Nagaya et al.~\cite{nagaya} and by Minakata and coworkers~\cite{nishio}. From theoretical viewpoint, we have to mention the work of
Manning~\cite{manning0}; his theory describes the individual ionic
activities of electrolyte-polyelectrolyte mixtures in semi-quantitative
agreement with experimental data. Recently Manning and Mohanty
extended the theory to solutions of charged oligomers~\cite{MM},
emphasizing the the importance of the ratio of the Debye screening
length $L_{\mathrm{D}}$ to the contour length of the polyion.

\begin{figure}[ht]
\begin{center}
\includegraphics[clip=true,width=0.48\textwidth]{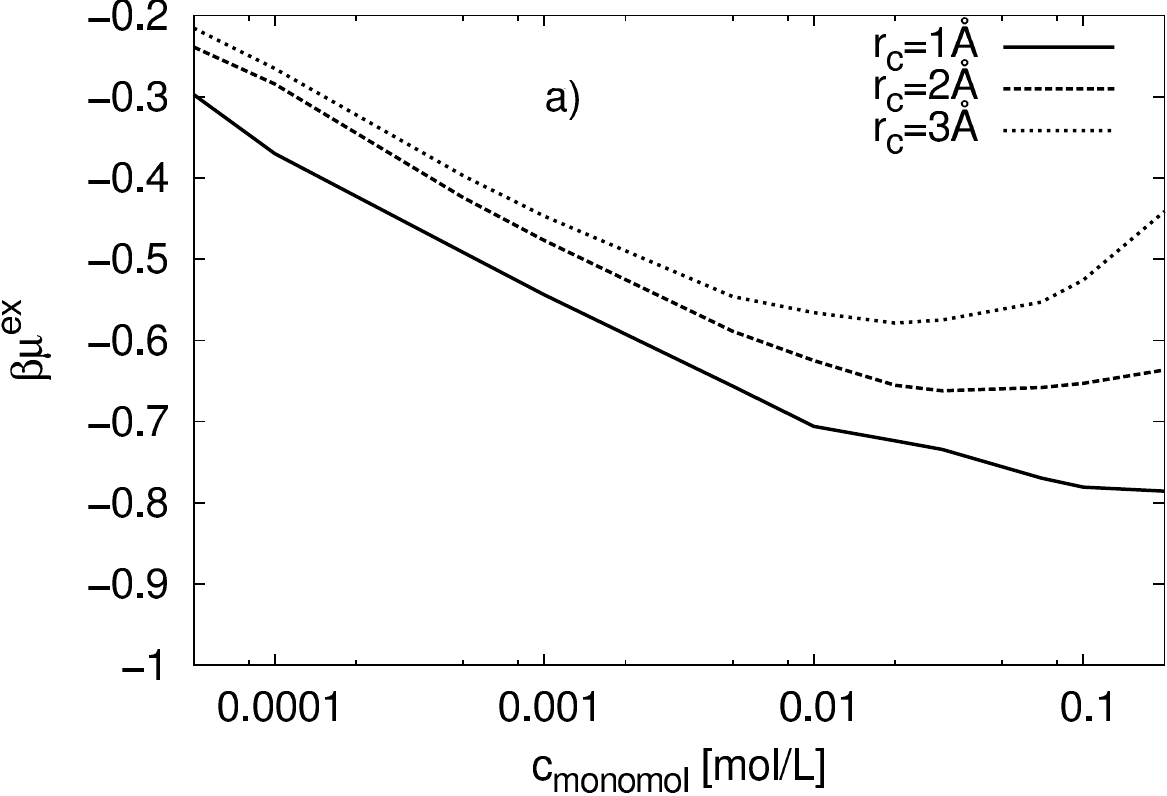}
\includegraphics[clip=true,width=0.48\textwidth]{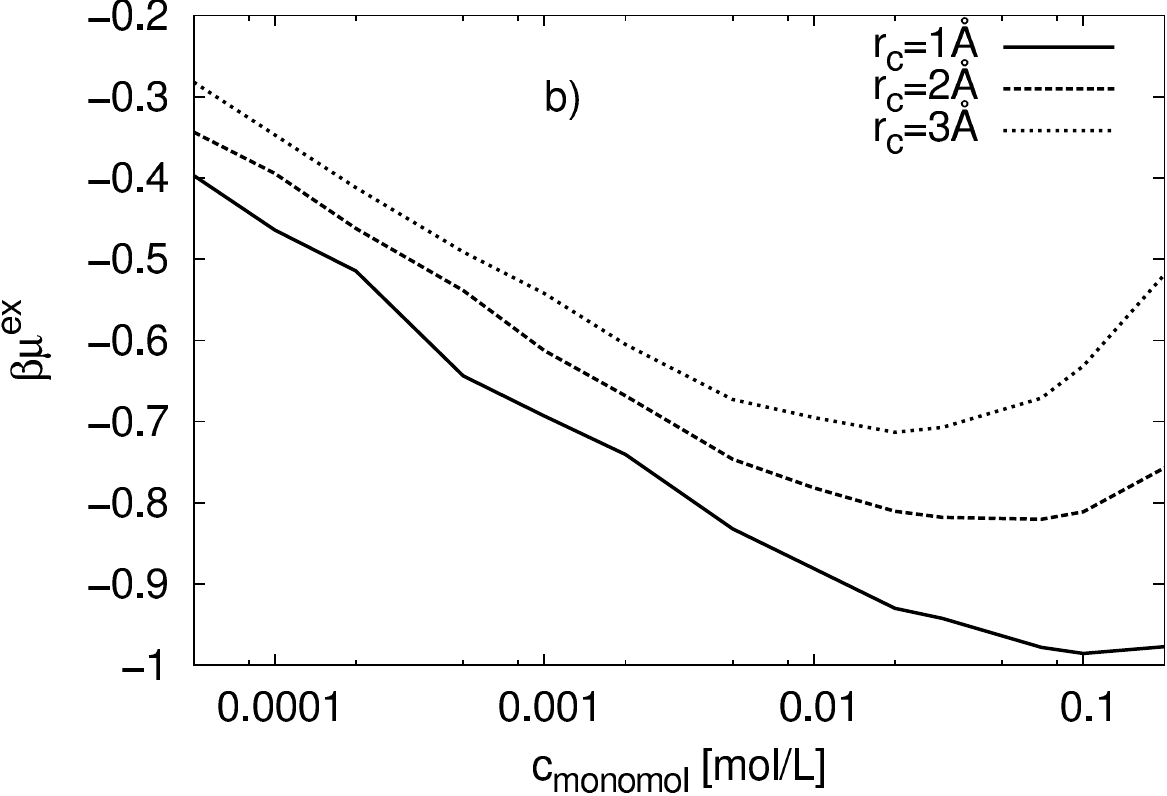}
\caption{Excess chemical potential of counterions for N=32 and for rigid (a)
and flexible chains~(b). }
\label{dil-H_}
\end{center}
\end{figure}

The results of our Monte Carlo simulation are shown in figure~10.  The
reduced excess chemical potential of counterions (logarithm of the
individual activity coefficient) exhibits quite a significant $r_{\mathrm{c}}$
dependence. The counterions of smaller size yield more negative
values of the excess chemical potential. This holds true for both rigid
(panel a) and flexible (panel b) oligoions. Another important
observation, not known from earlier studies, is that the $\beta \mu^{\mathrm{ex}}$
values are more negative in case of flexible oligoions. In other
words, it is less favorable to insert the counterion in solution
containing rigid oligoions. The effects are not small and should be
observed experimentally.
Finally, the last row of table~1 displays a dependence of $\beta \mu^{\mathrm{ex}}$
on chain length for flexible chains. A longer chain drives the
$\beta \mu^{\mathrm{ex}}$ toward more negative values.

\section{Conclusions}

New simulations and theoretical results were presented for two
models of polyelectrolyte solution. In one
case, the polyions were pictured as a rigid one-dimensional array of charged hard
spheres being totally rigid, while in the other example,
they were represented as a freely jointed (flexible) chain. The
results shown here strengthen and complement the conclusions
arrived at in some previous papers, most notably in~\cite{bizjak,antypov}.

Monte Carlo simulation was utilized to calculate the pair distribution
functions between various sites. The results for the
monomer-monomer, monomer -counterions, and
counterion-counterions distribution functions were obtained for the both
oligoion models, varying the radius of the counterions from 1.0~\AA ~to
3.0~\AA. The site-site pair distribution functions indicate significant
differences between the rigid and flexible model of the oligoion.  In
particular, the counterion-monomer pair distribution and
monomer-monomer distribution functions can differ quite
substantially if the counterions present in solution are small.

Among thermodynamic properties we choose to examine the enthalpy
of dilution and the activity coefficient of counterions.  These two
properties, taking into account the correlation between the activity and
osmotic coefficient, are widely used to characterize polyelectrolytes
in solutions. Experimental results for the enthalpy of dilution very
clearly reveal the ion-specific effects (see~\cite{cebasek} and the
references therein).  Aqueous solutions of $3,3$-ionenes (cationic
polyelectrolytes having quaternary ammonium groups on the
backbone) with fluoride (strong kosmotrope) counterions yield an
exothermic effect upon dilution in good agreement with theory and the
simulation results presented here.  On the other hand, the solutions
with chloride, bromide, and iodide counterions consume the heat
upon dilution (endothermic effect) in strong disagreement with the
theory and simulations based on the continuum solvent models. The
$3,3$-ionenes, studied in several papers~\cite{arh,Luksic1,Luksic2,Luksic3,cebasek}, have the charge density
parameter $\lambda=1.437$, which is not far from the value of the
present model, with $\lambda$ equal to 1.785. It is shown that the
parameters used to describe the polyelectrolyte models studied here
are not unrealistic. It is quite clear from the results presented in this
paper (cf. figure~9) that neither the effect of flexibility of polyions, nor the
size of the counterions (both in reasonable limits) can explain
qualitative differences between the continuum-solvent theories and
experimental results. As discussed elsewhere~\cite{cebasek}, the
ion-specific effects discussed above, are closely connected with the
hydration properties of the counterions and charged group on the
polyion.

The counterion activity coefficients in polyelectrolytes were studied in
many papers and the results were thoroughly analyzed by Wandrey and
coworkers~\cite{wandrey}. The experimental data confronted
with approximate theories. The analysis revealed important effects of
the polyelectrolyte concentration and the chain length of the polyion.
Measurements indicate that the activity coefficient of counterions
decreases with the increasing degree of polymerization. This finding
is consistent with our calculations. Our simulations, in agreement with
measurements, predict a substantial ion-specific effect; the smaller
is the ion the stronger is the ion binding~\cite{nagaya,arh}. In view of
theoretical nature of our study, where we can control the model
parameters, we have been able to examine the effect of the flexibility
of oligoion on the counterion activity coefficient. Such effects cannot
be easily studied experimentally. Our simulations show
that flexible oligoions (or polyions) attract the counterions more
strongly than the rigid ones. This finding supports the idea that
$\lambda_{\mathrm{eff}}
> \lambda$, should be used to successfully fit the experimental data.

In conclusion, the corrections due to the deviations from fully
extended polyion conformations are generally not large. This is
certainly true for highly charged polyelectrolytes examined here. The
situation, however, may be different in case of weakly charged
polyelectrolytes, or in the presence of large hydrophobic domains on the
polyion. In such situations,
transitions from the extended shape to the globular one are possible.

\section*{Acknowledgements}
This work was supported by the
Slovenian Research Agency through Physical Chemistry Research
Program 0103--0201.

\newpage
\ukrainianpart
\title{Кореляція між гнучкістю ланцюгоподібного поліелектроліту і
термодинамічними властивостями його розчину}

\author{Т. Саєвіч, Й. Решчіч, В. Влахи}

\address{Факультет хімії та хімічної технології, Університет
Любляни, Ашкерчева 5, SI--1000 Любляна}

\makeukrtitle
\begin{abstract}

Структурні та термодинамічні властивості модельного розчину,
що містить заряджені олігомери і еквівалентну кількість
протиіонів, вивчалися за допомогою канонічного моделювання Монте-Карло.
Олігомери представлені у вигляді (гнучких)
вільноз’єднаних ланцюжків або як лінійний (жорсткий) ланцюг заряджених твердих
сфер. Відповідно до примітивної моделі розчину електроліту,
протиіони моделюються як заряджені тверді сфери, а
розчинник враховується через діелектричні влас\-ти\-вос\-ті середовища.
Виявлено значні відмінності у парних функціях розподілу,
отриманих для жорсткої та гнучкої
моделей, але відмінності в термодинамічних властивостях,
таких як, ентальпія роз\-ве\-ден\-ня і надлишковий хімічний потенціал, є менш
значними. Отримані результати обговорюються з точки зору експериментальних даних
для водних розчинів поліелектролітів. Результати моделювання показують, що
відхилення від повністю витягнутої (стрижневої) конформації веде
до сильнішого зв'язування протиіонів. З іншого боку,
гнучкість полііонів, навіть у поєднанні з впливом розмірів іонів,
не може бути причиною якісних відмінностей між
теоретичними і експериментальними результатами для ентальпії розведення.

\keywords метод Монте-Карло, безсольовий розчин поліелектроліту, радіальні функції розподілу,
коефіцієнт активності, теплота розведення
\end{abstract}

\begin{thebibliography}{99}
\bibitem{alexandrowicz} Alexandrowicz~Z., Katchalsky~A.,
    J.~Polym. Sci., Part A: Polym. Chem.,  1963,  \textbf{1}, 3231--3260; \doi{10.1002/pol.1963.100011017}.

\bibitem{katchalsky} Katchalsky~A.,  Pure Appl. Chem., 1971, \textbf{26}, 327; \doi{10.1351/pac197126030327}.

\bibitem{dolar} Dolar~D. -- In: Polyelectrolytes, Selegnyi~E., ed. D. Reidel, Dordrecht, 1974, p.~97--113.

\bibitem{dautz} Dautzenberg~H., Jaeger~W., K\"{o}tz~J., Philipp~B.,
    Seidel~C., Stscherbina~D., Polyelectrolytes. Formation,
    Characterization and Application. Hanser, Munich, 1994.

\bibitem{schmitz} Schmitz~K.S. Macroions in Solution
    and Colloidal Suspension. VCH, New York, 1993.

\bibitem{forster} F\"{o}rster~S., Schmidt~M., Adv. Polym. Sci.: Physical
    Properties of Polymers, 1995, \textbf{120}, 51--133.

\bibitem{springer} Vlachy~V., Hribar-Lee~B.,  Re\v{s}\v{c}i\v{c}~J.,
    Kalyuzhnyi~Yu.V. -- In: Ionic Soft Matter: Modern Trends in
    Theory and Applications, NATO Science Series
    II: Mathematics, Physics and Chemistry, Henderson~D., Holovko~M. Trokhymchuk~A. (eds.). Springer, 2005.

\bibitem{dobrynin1} Dobrynin~A.V., Curr. Opin. Colloid Interface Sci., 2008, \textbf{13}, 376--388; \doi{10.1016/j.cocis.2008.03.006}.

\bibitem{wandrey} Wandrey~C., Hunkeler~D., Wendler~U., Jaeger~W., Macromolecules, 2000, \textbf{33}, 7136--7143; \\  \doi{10.1021/ma991763d}.

\bibitem{blaul} Blaul~J., Wittemann~M., Ballauff~M., Rehahn~M., J.~Phys. Chem.~B,  2000, \textbf{104}, 7077--7081; \\ \doi{10.1021/jp001468r}.

\bibitem{Holm2004} Holm~C., Rehahn~M., Opperman~W., Ballauff~M., Adv. Polym. Sci., 2004, \textbf{166}, 1--27; \\ \doi{10.1007/b11347}.

\bibitem{jerman}Jerman B., Breznik M., Kogej K.,
    Paoletti S., J.~Phys.~Chem. B, 2007, \textbf{111}, 8435--8443; \\ \doi{10.1021/jp0676080}.

\bibitem{manning0} Manning~G.S.,  J.~Chem.~Phys., 1969, \textbf{51}, 924; \doi{10.1063/1.1672157}.

\bibitem{manning1} Manning~G.S.,  J.~Chem.~Phys., 1969, \textbf{51}, 934; \doi{10.1063/1.1672158}.

\bibitem{vesnaver0} Vesnaver~G., Dolar~D., Eur. Polym. J., 1976, \textbf{12}, 129--132; \doi{10.1016/0014-3057(76)90039-2}.

\bibitem{arh} Arh~K., Pohar~C., Vlachy~V., J. Phys. Chem. B,
     2002, \textbf{106}, 9967--9973; \doi{10.1021/jp025858k}.

\bibitem{lipar} Lipar~I., Zalar~P., Pohar~C., Vlachy~V.,
     J. Phys. Chem. B,  2007,  \textbf{111}, 10030--10136;\\ \doi{10.1021/jp073641q}.

\bibitem{nagaya} Nagaya~J., Minakata~A., Tanioka~A., Langmuir,  1999, \textbf{15}, 4129--4134; \doi{10.1021/la981190i}.

\bibitem{Luksic1} Luk\v{s}i\v{c}~M., Buchner~R., Hribar-Lee~B.,
    Vlachy~V.,  Macromolecules,  2009,  \textbf{42}, 4337--4342; \\ \doi{10.1021/ma900097c}.

\bibitem{Luksic2} Luk\v{s}i\v{c}~M., Buchner~R., Hribar-Lee~B., Vlachy~V.,  Phys. Chem. Chem. Phys.,  2009,
     \textbf{11}, 10053--10058; \\ \doi{10.1039/B914938B}.

\bibitem{Luksic3}  Luk\v{s}i\v{c}~M., Hribar-Lee~B., Vlachy~V.,
     J. Phys. Chem. B,  2010, \textbf{114}, 10401--10408; \\ \doi{10.1021/jp105301m}.

\bibitem{kremer2} Stevens~M.J., Kremer~K., Phys. Rev. Lett.,  1993, \textbf{71}, 2228--2231; \doi{10.1103/PhysRevLett.71.2228}.

\bibitem{kremer3} Stevens~M.J., Kremer~K., J. Chem. Phys.,  1995,  \textbf{103}, 1669--1690; \doi{10.1063/1.470698}.

\bibitem{osm} Chang~R., Yehtiraj~A.,  Macromolecules, 2005,  \textbf{38}, 607--616; \doi{10.1021/ma0486952}.

\bibitem{bizjak} Bizjak~A., Re\v{s}\v{c}i\v{c}~J., Kalyuzhnyi~Yu.V., Vlachy~V., J. Phys. Chem. B,  2006, \textbf{125}, 214907; \\ \doi{10.1063/1.2401606}.

\bibitem{antypov} Antypov~D., Holm~C., Macromolecules,  2007, \textbf{40}, 731--738; \doi{10.1021/ma062179p}.

\bibitem{VD} Vlachy~V., Dolar~D., J. Chem. Phys., 1982, \textbf{76}, 2010--2014; \doi{10.1063/1.443174}.

\bibitem{BD} Bratko~D., Dolar~D.,  J. Chem. Phys., 1984,  \textbf{80}, 5782--5789; \doi{10.1063/1.446601}.

\bibitem{Carrillo} Carrillo~J.-M.Y., Dobrynin~A., J. Phys. Chem. B, 2010, \textbf{114}, 9391--9399; \doi{10.1021/jp101978k}.

\bibitem{Aseyev} Aseyev~V.O., Klenin~S.I., Tenhu~H., Grillo~I., Geissler~E.,  Macromolecules,  2001,  \textbf{34}, 3706--3709; \\ \doi{10.1021/ma0016331}.

\bibitem{Essafi} Essafi~W., Spiteri~M.N., Williams~C., Boue~F., Macromolecules,  2009, \textbf{42}, 9568--9580; \\ \doi{10.1021/ma9003874}.

\bibitem{vesnaver} Vesnaver G., Rude\v{z}~M., Pohar~C.,
    \v{S}kerjanc~J.,  J. Phys. Chem.,  1984,  \textbf{88}, 2411--2418; \\ \doi{10.1021/j150655a046}.

\bibitem{cebasek} \v{C}ebasek~S., Luk\v si\v c~M., Pohar~C.,
    Vlachy~V.,  J. Chem. Eng. Data,  2011, \textbf{56}, 1282; \doi{10.1021/je101136a}

\bibitem{molsim} Linse~P., Molsim 3.6.7. Lund University,
    Sweden, 2004.

\bibitem{jesus1} Pi\~{n}ero~J., Bhuiyan~B.L., Re\v{s}\v{c}i\v{c}~J., Vlachy~V., Acta Chim. Slov.,
     2006, \textbf{53}, 316--323.

\bibitem{jesus2} Pi\~{n}ero~J., Bhuiyan~B.L., Re\v{s}\v{c}i\v{c}~J., Vlachy~V.,  J. Chem. Phys.,
     2007, \textbf{127}, 104904; \\ \doi{10.1063/1.2768963}.

\bibitem{jesus2a} Pi\~{n}ero~J., Bhuiyan~B.L., Re\v{s}\v{c}i\v{c}~J., Vlachy~V.,  J. Chem. Phys.,
     2008,  \textbf{128}, 119901; \\ \doi{10.1063/1.2841082}.

\bibitem{jesus3} Pi\~{n}ero~J., Bhuiyan~B.L., Re\v{s}\v{c}i\v{c}~J., Vlachy~V., J. Chem. Phys.,
     2008,  \textbf{128}, 214904; \\ \doi{10.1063/1.2919134}.

\bibitem{morawetz1} Morawetz~H.,  Acc. Chem. Res., 1970, \textbf{3}, 354--360; \doi{10.1021/ar50034a005}.

\bibitem{morawetz2}  Morawetz~H., J. Polym. Sci., Part B: Polym. Phys.,  2002, \textbf{40}, 1080--1086; \doi{10.1002/polb.10167}.

\bibitem{RVH} Re\v{s}\v{c}i\v{c}~J., Vlachy~V., Haymet~A.D., J. Am. Chem. Soc.,  1990,  \textbf{112}, 3398--3401; \\   \doi{10.1021/ja00165a022}.

\bibitem{Boyd} Boyd~G.E., Wilson~D.P., J. Phys. Chem., 1976,  \textbf{80}, 805--808; \doi{10.1021/j100549a006}.

\bibitem{cation} Keller~M., Lichtentaler~R.N., Heintz~A.,
     Ber. Bunsen Ges. Phys. Chem., 1996,  \textbf{100}, 776--779.

\bibitem{vlachy2008} Vlachy~V., Pure \& Applied Chemistry, 2008,  \textbf{80}, 1253--1266; \doi{10.1351/pac200880061253}.

\bibitem{pohar} Arh~K., Pohar~C.,  Acta Chim. Slov., 2001,  \textbf{48}, 385--394.

\bibitem{frenkel} Frenkel~D., Smit~B., Understanding
   Molecular Simulation, From Algorithms to Applications. Academic Press, 1996.

\bibitem{gueron} Gueron M., Weisbuch~G.,  J. Phys. Chem., 1979, \textbf{83}, 1991--1998; \doi{10.1021/j100478a013}.

\bibitem{nagasawa} Nagasawa~M., Kagawa~I., J. Polym. Sci., 1957, \textbf{25}, 61--76; \doi{10.1002/pol.1957.1202510806}.

\bibitem{oman} Oman~S., Dolar~D.,  Z. Phys. Chem. N.F., 1967, \textbf{56}, 1--12.

\bibitem{joshi} Joshi~Y.M., Kwak~J.C.T,  J. Phys. Chem., 1979, \textbf{83}, 1978--1983; \doi{10.1021/j100478a011}.

\bibitem{nishio} Nishio~T., Minakata~A.,  Langmuir,  1999, \textbf{15}, 4123--4128; \doi{10.1021/la981188r}.

\bibitem{MM} Manning~G.S., Mohanty~U., Physica A, 1997, \textbf{247}, 196--204; \doi{10.1016/S0378-4371(97)00413-5}.


\end{thebibliography}
\end{document}